%% file: main.tex
\documentclass[manuscript,screen]{acmart}

\newcommand{\cmk}{\textcolor{rq}{\ding{51}}}
\newcommand{\xmk}{\textcolor{rd}{\ding{55}}}

\AtBeginDocument{%
  }

\setcopyright{acmlicensed}
\copyrightyear{2026}
\acmYear{2026}
\acmDOI{XXXXXXX.XXXXXXX}

\acmJournal{TOSEM}
\acmVolume{0}
\acmNumber{0}
\acmArticle{0}
\acmMonth{0}

\usepackage{tabularx}  
\usepackage{booktabs}  
\usepackage{threeparttable}  

\usepackage{pifont}

\usepackage{pgfplots}
\pgfplotsset{compat=1.18}
\usepackage{siunitx}
\usepackage[export]{adjustbox}
\usepackage{subcaption}
\usepackage[inkscapearea=page]{svg}
\usepackage{pgf-pie}

\usepackage{tikz}
\usetikzlibrary{shapes,shadows,arrows}
\tikzstyle{rqbox} = [draw=color_graph2, fill=color_graph3, very thick,
    rectangle, rounded corners, inner sep=10pt, inner ysep=12pt]
\tikzstyle{titlerq} =[fill=color_graph2, draw=color_graph2,  rounded corners, inner sep=4pt]
\definecolor{rq}{HTML}{2a918c}
\definecolor{crq}{HTML}{ededed}
\definecolor{rd}{HTML}{FF4242} 
\definecolor{nd}{HTML}{121821}
\definecolor{bl}{HTML}{019fe2}
\definecolor{yl}{HTML}{efcf42}

\definecolor{crq2}{HTML}{dee2e6}
\definecolor{crq4}{HTML}{425168}
\definecolor{crq3}{HTML}{535662}

\definecolor{color_graph}{HTML}{6c757d}
\definecolor{color_graph1}{HTML}{adb5bd}
\definecolor{color_graph2}{HTML}{dee2e6}
\definecolor{color_graph3}{HTML}{f8f9fa}

\usepackage[export]{adjustbox}
\usepackage[edges]{forest}

\usepackage{pgfplots}
\usepgfplotslibrary{groupplots}
\usepackage{makecell}
\usepackage{multirow}

\definecolor{blueGray1}{RGB}{142, 158, 171}
\definecolor{blueGray2}{RGB}{112, 132, 148}
\definecolor{blueGray3}{RGB}{82, 102, 118}
\definecolor{orange1}{RGB}{255, 180, 120}
\definecolor{orange2}{RGB}{255, 140, 60}
\definecolor{orange3}{RGB}{235, 120, 40}
\definecolor{green1}{RGB}{150, 200, 150}
\definecolor{green2}{RGB}{120, 180, 120}
\definecolor{red1}{RGB}{255, 150, 150}
\definecolor{red2}{RGB}{255, 120, 120}
\definecolor{redGray1}{RGB}{226, 176, 168}
\definecolor{redGray2}{RGB}{193, 109, 97}

\definecolor{rqa}{HTML}{344465}
\definecolor{rqb}{HTML}{ccd0d8}

\usepackage[most]{tcolorbox}
\newtcolorbox{rqanswer}[1]{
  colback=rqb, colframe=rqa,
  title={Answer to #1}, fonttitle=\bfseries,
  boxrule=0.5pt, left=4pt, right=4pt, top=3pt, bottom=3pt,
  breakable
}

\definecolor{rqa}{HTML}{344465}
\definecolor{rqb}{HTML}{ccd0d8}
\usepackage[most]{tcolorbox}
\newtcolorbox{answerbox}{
  enhanced,
  breakable,
  colback=rqb,
  colframe=rqb,
  boxrule=0pt,
  frame hidden,
  borderline west={3pt}{0pt}{rqa},
  sharp corners,
  left=8pt, right=8pt, top=6pt, bottom=6pt
}

\begin{document}

\title{Software Security in Software-Defined Networking: A Systematic Literature Review}


\author{Moustapha Awwalou Diouf}
\orcid{0009-0000-2063-5175}
\affiliation{%
  \institution{SnT, University of Luxembourg}
  \country{Luxembourg}
}
\email{moustapha.diouf@uni.lu}

\author{Samuel Ouya}
\orcid{0000-0002-8665-8131}
\affiliation{%
  \institution{Cheikh H. KANE Digital University}
  \country{Senegal}}
\email{samuel.ouya@unchk.edu.sn}

\author{Jacques KLEIN}
\orcid{0000-0003-4052-475X}
\affiliation{%
  \institution{SnT, University of Luxembourg}
  \country{Luxembourg}
}
\email{jacques.klein@uni.lu}

\author{Tegawendé F. Bissyandé}
\orcid{0000-0001-7270-9869}
\affiliation{%
 \institution{SnT, University of Luxembourg}
 \country{Luxembourg}}
 \email{tegawende.bissyande@uni.lu}





\renewcommand{\shortauthors}{Diouf et al.}

\begin{abstract}
\input{Sections/0-abstract}
\end{abstract}




\maketitle

\section{Introduction}
\input{Sections/1-introduction}

\section{Background}\label{sec:bckg}
\input{Sections/2-background}

\section{Methodology}\label{sec:meth}
\input{Sections/3-methodology}

\section{Results}\label{sec:rslt}
\input{Sections/4-result}

\section{Challenges and Opportunities}\label{sec:challenges}
\input{Sections/9-challenges}

\section{Threats To Validity}\label{sec:ttv}
\input{Sections/6-threatsToValidity}

\section{Related Work}\label{sec:rw}
\input{Sections/7-relatedWork}

\section{Conclusion}\label{sec:concl}
\input{Sections/8-conclusion}

\bibliographystyle{ACM-Reference-Format}
\bibliography{sample-base}

\end{document}

%% file: Sections/0-abstract.tex
Software-defined networking (SDN) separates the control plane from the data plane and exposes the network through software applications and open APIs. The same programmability that drove its adoption also turned the network into a large body of software, and that software can be vulnerable. Knowing where and how SDN software fails is a prerequisite for defending it, yet the literature offers no consolidated account of the problem. We address this gap with a systematic literature review of 122 primary studies published between 2012 and 2026 on the security of the software that makes up SDN. We study how each software component becomes vulnerable, how attackers exploit it, which testing and analysis techniques expose its defects, and how the field has evolved. The review yields a taxonomy of vulnerabilities and attack vectors organized by SDN software component, a synthesis of the methods used to find software defects in each component, and a set of open problems that mark the most promising directions for future work. Earlier surveys treat SDN security as a networking problem; ours is, to our knowledge, the first to treat SDN components as software artifacts whose code, logic, and interactions can be defective. Our artifacts are available at \url{https://github.com/mad975/SDNSoftwareSecurity_SLR}.

%% file: Sections/1-introduction.tex
Our information-driven economy rests on data networks, but traditional network architectures cannot keep pace with the demands of cloud computing, the Internet of Things, ubiquitous mobile computing, and big-data analytics~\cite{GuoYang01}. Building networks from purpose-specific, fixed-function hardware such as routers, switches, firewalls, and load balancers no longer matches the economics of modern virtualized computing~\cite{HAUSER2023103561}. The result is long development cycles for new network services and high capital and operational costs~\cite{nunes2014survey}.

Software-defined networking (SDN) emerged in response. SDN decouples the control plane, which holds the network's control logic, from the data plane, which forwards packets, and replaces fixed-function devices with a programmable, software-driven network. A logically centralized controller orchestrates the network according to administrator-defined policy. Applications run on top of the controller and tell it how to operate the network, forming an application layer, and the controller exposes its functions to those applications through application programming interfaces (APIs). This flexibility has pushed SDN into data centers, 5G systems, and other large-scale infrastructure~\cite{KHORSANDROO2021107981}, a trend reinforced by the continued growth of the user base the network must serve, which reached about 6 billion people in 2025~\cite{ITU2025}; North America alone accounts for more than 35\% of the global SDN market~\cite{GMI2395}.

This same programmability reshapes the network's security. The logically centralized controller becomes a single point of failure, monitoring grows harder, and the programmable interfaces between planes open new attack surfaces. Most SDN security research, however, inherits its agenda from traditional networking and concentrates on network-level threats such as intrusion detection~\cite{8024623, 10.1145/3309194.3309199, 9187858, sultana2019survey, 9919834, 7838212}, defense against DoS and DDoS attacks~\cite{bawany2017ddos, 7524500, 7289347, 8030502, 10489935}, and related threats carried over from legacy networks~\cite{8566268, 10.1145/2876019.2876028, 9547905}. The security of the software that constitutes SDN has received far less attention, even though software defects are a leading cause of outages in large production networks. Software bugs account for roughly a third of high-impact failures at Google~\cite{10.1145/2934872.2934891}, and about 60\% of the outages studied at Facebook were software related~\cite{10.1145/3230543.3230546}. The few prior efforts that look at SDN software remain narrow. One examines controller bugs for reliability rather than security~\cite{9505089}; another examines ways to improve deployment security~\cite{10226193}. Both leave the wider software ecosystem, namely third-party applications, APIs, and the data plane, largely unexamined. To our knowledge, no systematic review has comprehensively examined the security of SDN software.

In this article, we present a systematic literature review (SLR) of research that targets the security of the software integral to SDN. By establishing the state of the art, identifying its strengths, weaknesses, and gaps, and outlining future directions, we give researchers and practitioners a guide to the intersection of SDN and software-security engineering. This work makes the following contributions.
\begin{itemize}
\item We conduct a systematic review of 122 studies (2012--2026) that map SDN software security across its four software components, namely the controller, the applications, the interfaces, and the data plane, together with the security activities applied to each. This software-centric mapping distinguishes our work from prior reviews centered on network-level security.
\item We analyze how SDN software components are tested and analyzed for security defects, characterizing the testing and program-analysis techniques the literature employs and where they are applied.
\item We derive a taxonomy of vulnerabilities and attack vectors specific to SDN software components, organized by their location in the SDN architecture, and we map each attack to the vulnerability it exploits and to the software-level defenses reported against it.
\item We synthesize the main open security challenges of SDN software components and set out concrete directions for future research.
\end{itemize}

The remainder of this article is organized as follows. Section~\ref{sec:bckg} introduces SDN and the security of its software components. Section~\ref{sec:meth} describes the methodology, from the research questions and the search to study selection, snowballing, and data extraction. Section~\ref{sec:rslt} presents the results, with an overview of the corpus followed by the answers to RQ1 to RQ4. Section~\ref{sec:challenges} derives challenges and opportunities from these results. Section~\ref{sec:ttv} discusses threats to validity, Section~\ref{sec:rw} compares this review with prior SDN-security reviews, and Section~\ref{sec:concl} concludes.

%% file: Sections/2-background.tex
\subsection{What is software-defined networking?}\label{sec:wsdn}
\begin{figure}[h!]
  \centering
  \includegraphics[width=0.8\linewidth]{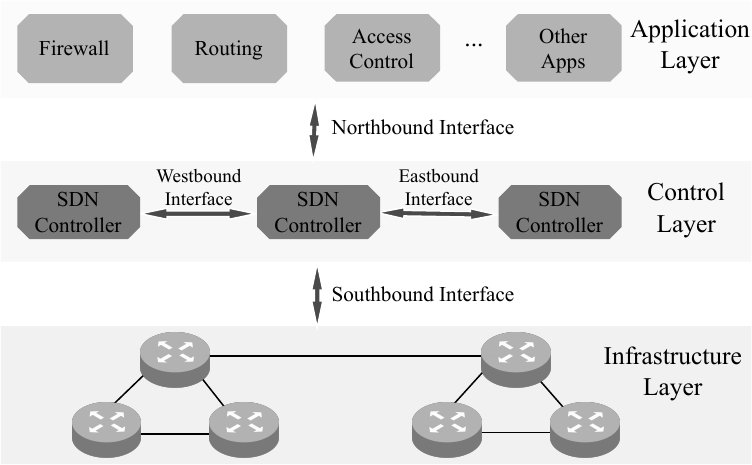}
  \caption{SDN architecture.}
  \label{fig_sdn}
\end{figure}

SDN offers flexibility, abstraction, programmability, and virtualization to overcome the limitations of traditional network architectures~\cite{singh2020detection}. The Open Networking Foundation~\cite{onf2024} conceptualizes the SDN architecture as three layers, illustrated in Figure~\ref{fig_sdn}: the infrastructure layer corresponds to the data plane, while the control functionality above it is separated into a control layer and an application layer to improve programmability.

\noindent
\textbf{Control layer.} This layer contains one or more controllers. The complex control logic is implemented in the controller, often called the brain of SDN. The control layer offers the flexibility to add new functionality through programming interfaces~\cite{9349440}.

\noindent
\textbf{Application layer.} This layer contains applications that implement network services such as firewalls, load balancing, and intrusion detection. An SDN application is a software program deployed on the controller. Such applications communicate with the controller through a northbound interface (e.g., REST) according to their network requirements~\cite{10.1145/3453648}. The control layer presents an abstract view of the physical elements to the application layer, and the applications define the decision-making logic that the control layer enforces.

\noindent
\textbf{Infrastructure layer.} This layer comprises the switches and routers that forward packets according to controller instructions. These devices communicate with the controller through a southbound interface, such as the OpenFlow protocol~\cite{liatifis2023advancing}. The data plane is not limited to fixed-function hardware. With P4~\cite{bosshart2014p4} or eBPF~\cite{hoiland2018xdp}, operators can reprogram how packets are processed; a data plane whose forwarding pipeline is reconfigurable in this way is called a programmable data plane. Virtual switches such as Open vSwitch, by contrast, implement forwarding in software~\cite{10.1145/3185467.3185468}.

\noindent
\textbf{Application Programming Interfaces (APIs).} APIs enable communication between the SDN layers. The southbound interface connects the data plane and the control plane; NETCONF~\cite{8657331} and OpenFlow~\cite{liatifis2023advancing} are two common examples. The northbound interface connects the controller and the SDN applications, and the east-west interface lets controllers in different SDN domains share information~\cite{9349440}. Northbound and east-west interfaces are not yet standardized.

\subsection{Security of SDN software components}\label{sec:ssdnc}
Each layer of the SDN architecture is realized as software, and software can contain exploitable defects. The controller is a network operating system that hosts the applications, so faulty or malicious controller code can cause a loss of network control~\cite{10.1145/2660267.2660353}. Not every defect lives inside a single component. A growing class of logic bugs arises from the way components interact, and conventional analysis tools, which reason about one component at a time, tend to miss them~\cite{Ujcich2020}. Reasoning about SDN security therefore means reasoning about both the components and their interactions.

SDN control is event-driven, which opens a distinctive attack surface. When a packet does not match any installed rule, the switch forwards it to the controller as an event, and the controller dispatches that event to its applications~\cite{Lee2017}. An attacker who controls a host can therefore craft packets that the controller will act on, and through them escalate privileges or manipulate flow rules without ever compromising the controller directly~\cite{Ujcich2020}. The data plane thus becomes a channel through which untrusted input reaches privileged control logic. Attacks that use this channel are called cross-plane attacks~\cite{Ujcich2020}. Cross-plane analysis follows how events are used and passed between SDN components and how their actions propagate within the control plane to have data-plane effects~\cite{Ujcich2020}.

The application layer compounds this risk. Applications reach the controller through the northbound interface, where third-party and untrusted code operates on shared control-plane state~\cite{10.1145/3243734.3243759}. In cross-app poisoning, an unprivileged application manipulates this shared state so that a privileged application takes actions on its behalf~\cite{10.1145/3243734.3243759}. A single faulty or malicious application can crash the controller or seize control of every switch in the network~\cite{10.1145/2660267.2660353, 10.1145/2342441.2342466}. Because control is logically centralized, the blast radius of such a compromise is the entire network rather than one device. The data plane is itself a software attack surface, since programmable forwarding with P4 or eBPF and software switches such as Open vSwitch can contain exploitable bugs~\cite{10.1109/sp61157.2025.00194, 10.1145/3185467.3185468}. The security of an SDN deployment, then, is the combined security of its applications, interfaces, controller, and data plane.

%% file: Sections/3-methodology.tex
\begin{figure}[h!]
  \centering
  \includegraphics[width=\linewidth]{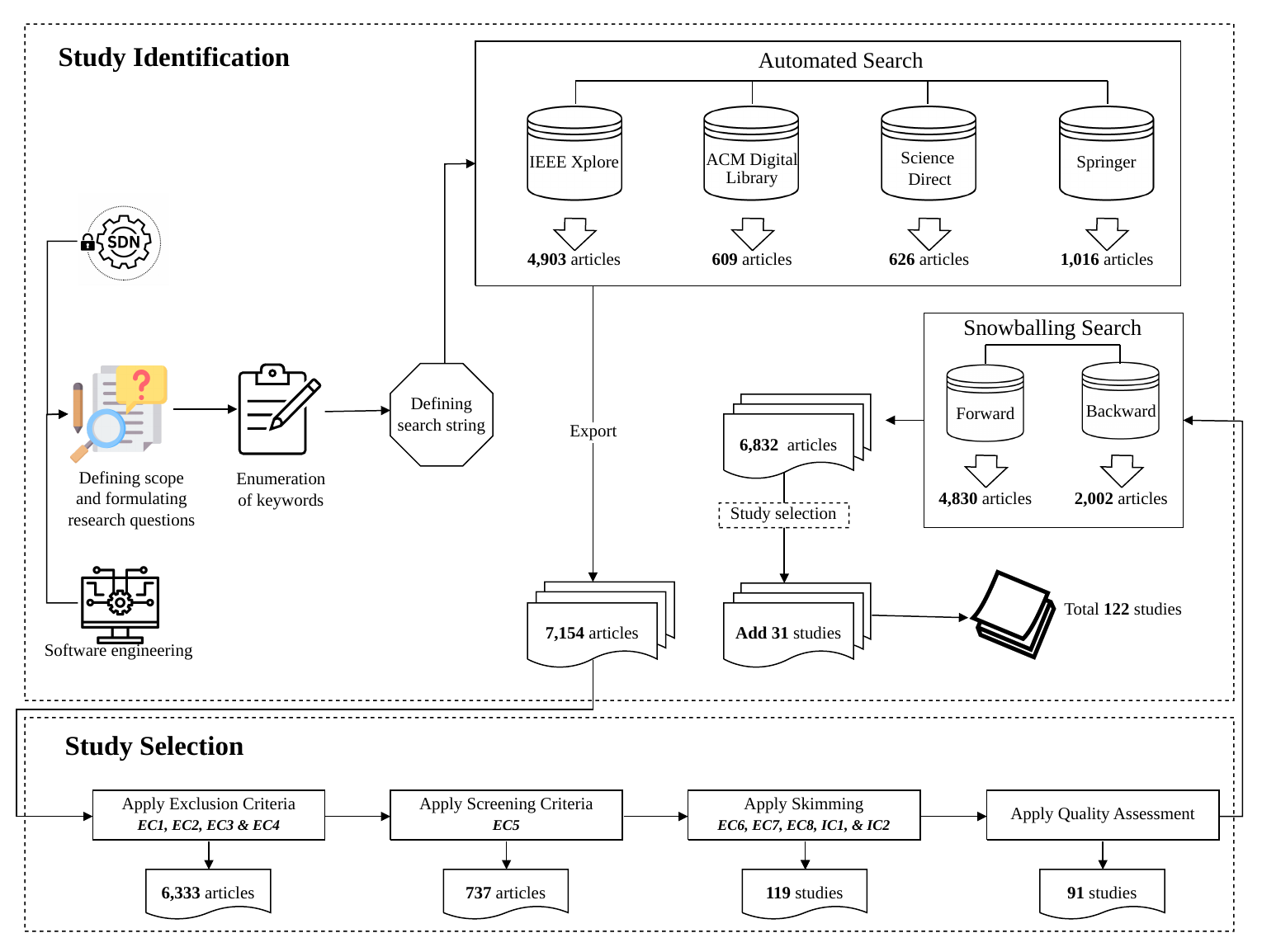}
  \caption{Summary of our SLR process.}
  \label{fig_pr_slr}
\end{figure}

We follow the systematic literature review (SLR) methodology, an established approach in software-engineering research~\cite{LI201767, 10.1145/3450288, 6035727}. Adopting the guidelines of Kitchenham and Charters~\cite{kitchenham2007guidelines}, we organize the review into three phases. We plan the review in Sections~\ref{sec:rqs} and~\ref{sec:crq}, conduct it in Sections~\ref{sec:ss} to~\ref{sec:dea}, and report its results from Section~\ref{sec:rslt} onward. Figure~\ref{fig_pr_slr} gives an overview of the process, which the following sections describe in detail.

\subsection{Research Question}\label{sec:rqs}
We formulate four research questions to guide our SLR on software security in SDN.
\begin{itemize}
\item \textbf{RQ1: What are the current trends in the literature on software security in SDN?} This question identifies the main themes and emerging patterns in SDN software security and traces how the field has evolved.
\item \textbf{RQ2: What methods are used to test and analyze security vulnerabilities in SDN software components?} This question examines the techniques the literature uses to assess SDN software, uncovering the testing and analysis strategies most commonly employed.
\item \textbf{RQ3: How can SDN software vulnerabilities be classified?} This question identifies the vulnerabilities that affect SDN software components and organizes them into a taxonomy.
\item \textbf{RQ4: What attack patterns target SDN software vulnerabilities, how do they exploit these weaknesses, and what defenses mitigate them?} This question maps each attack to the vulnerability it exploits and to the software-level defenses reported against it, to support the prioritization of defensive strategies.
\end{itemize}

\subsection{Search query construction}\label{sec:crq}
We construct our search query from keywords derived from our research questions, organized into three categories as shown in Table~\ref{tab:sk}. Within each category, keywords are combined with the $OR$ operator; the three categories are then combined with $AND$ (\textit{finalString} = $C_1 \land C_2 \land C_3$).

\begin{table}[ht!]
\caption{Search Keywords}
\label{tab:sk}
\centering
\begin{tabularx}{0.9\textwidth}{lX}
\toprule
\textbf{Category} & \textbf{Keywords} \\ 
\midrule
1 & Software-Defined Network*, Software Defined Network*, SDN* \\ 

2  & Controller*, Application*, OpenFlow, Southbound, Northbound, Api*, Interface* \\ 

3  & Malware, Malicious, Attack*, Exploit*, Detect*, Test*, Analy*, Fault*, Bug*, Threat*, issue*, Vulnerab*, Assess*, Verification*, Taxonom* \\ 
\bottomrule
\end{tabularx}
\par\smallskip
\begin{minipage}{0.9\textwidth}
\footnotesize Note: * indicates wildcard search terms (e.g., ``Vulnerab*'' matches Vulnerability, Vulnerabilities).
\end{minipage}
\end{table}

To maximize the retrieval of relevant publications, we target the four most popular repositories: IEEE Xplore\footnote{https://ieeexplore.ieee.org/Xplore/home.jsp}, ACM Digital Library\footnote{https://dl.acm.org/}, ScienceDirect\footnote{https://www.sciencedirect.com/}, and SpringerLink\footnote{https://www.springer.com/}, using the advanced search of each. We applied the same core search string to all four databases, adjusting it to each one's syntax and constraints; the full per-database strings are in our replication package\footnote{\url{https://github.com/mad975/SDNSoftwareSecurity_SLR}}. IEEE Xplore and the ACM Digital Library were searched on title and abstract. SpringerLink and ScienceDirect were searched on full text, then restricted to records matching the query in their title or abstract. SDN emerged as a distinct research field with the founding of the Open Networking Foundation in 2011~\cite{SCHALLER2017197}. We therefore searched publications from 2011 to September 2026; the earliest study in our corpus dates from 2012. This search yielded \texttt{7,154} articles: \texttt{4,903} from IEEE Xplore, \texttt{609} from ACM Digital Library, \texttt{626} from ScienceDirect, and \texttt{1,016} from SpringerLink.

\subsection{Study Selection} \label{sec:ss}
\noindent
\textbf{Inclusion and Exclusion Criteria. } The \texttt{7,154} publications obtained are potentially related to our research. However, using broad keywords to maximize coverage retrieves many irrelevant publications. To ensure a focused and reliable set of primary studies for our SLR, we apply a three-phase filtering method to assess relevance according to the inclusion and exclusion criteria detailed in Table~\ref{tab:criteria}.

\begin{table}[ht!]
  \caption{Inclusion and Exclusion Criteria}
  \label{tab:criteria}
  \centering
  \begin{tabular}{cp{0.8\textwidth}}
    \toprule
    \multicolumn{2}{l}{\textbf{Inclusion Criteria}} \\
    \midrule
    \textcolor{rq}{\ding{51}} & Studies whose primary contribution addresses the security of SDN software components (controller, applications, interfaces, or the data plane) through software engineering methods or software-level mechanisms. \footnotesize{\textit{(IC1)}} \\

    \textcolor{rq}{\ding{51}} & Studies with accessible full text. \footnotesize{\textit{(IC2)}} \\
    \midrule
    \multicolumn{2}{l}{\textbf{Exclusion Criteria}} \\
    \midrule
    \textcolor{rd}{\ding{55}} & Non-English written literature. \footnotesize{\textit{(EC1)}} \\
    \textcolor{rd}{\ding{55}} & Studies with fewer than 5 pages. \footnotesize{\textit{(EC2)}} \\
    \textcolor{rd}{\ding{55}} & Studies belonging to books, theses, or keynotes. \footnotesize{\textit{(EC3)}} \\
    \textcolor{rd}{\ding{55}} & Duplicate studies or similar studies with the same title from the same authors. \footnotesize{\textit{(EC4)}} \\
    \textcolor{rd}{\ding{55}} & Studies where title and abstract show no connection to software security in SDN. \footnotesize{\textit{(EC5)}} \\
    \textcolor{rd}{\ding{55}} & Studies that only mention software security in SDN in discussions or as future work. \footnotesize{\textit{(EC6)}} \\
    \textcolor{rd}{\ding{55}} & Studies whose primary contribution addresses classical network security in SDN deployments rather than the security of SDN software components as software artifacts. \footnotesize{\textit{(EC7)}} \\
    \textcolor{rd}{\ding{55}} & Not available in full text. \footnotesize{\textit{(EC8)}} \\
    \bottomrule
  \end{tabular}
\end{table}

In the first phase, we applied exclusion criteria EC1, EC2, EC3, and EC4 to remove publications in languages other than English, short articles of fewer than five pages, non-peer-reviewed content such as books, theses, and keynotes, and duplicates, which reduced the set from \texttt{7,154} to \texttt{6,333} articles. In the second phase, we screened the title and abstract of each remaining article against EC5 to remove studies unrelated to software security in SDN, which left \texttt{737} candidates. In the third phase, we read the full text of every candidate and applied exclusion criteria EC6, EC7, and EC8 together with inclusion criteria IC1 and IC2, retaining only studies that substantially addressed software security in SDN from a software-engineering perspective. This phase yielded \texttt{119} candidate studies for quality assessment. Two authors independently screened the candidates in each phase, and disagreements were resolved through discussion and consensus.

\noindent
\textbf{Quality Assessment. }To guard against bias from low-quality studies, we defined four Quality Assessment Criteria (QAC), shown in Table~\ref{tab:qac}, that evaluate the methodological clarity, the validity of the results, and the contribution of each study. We scored each of the \texttt{119} candidates manually on every criterion using a four-point scale from 0 to 3 (poor, average, good, excellent), giving a maximum of 12 points. We retained studies scoring at least 8 out of 12, equivalent to an average of 2 (good) per criterion, which left \texttt{91} studies.

\begin{table}[ht!]
  \caption{Checklist of Quality Assessment Criteria (QAC) for SDN software}
  \label{tab:qac}
  \begin{tabular}{ll}
    \toprule
        \textbf{ID} & \textbf{QAC} \\ 
        \midrule
        QAC1 &  Are the research objectives clearly defined?  \\ 
        QAC2 &  Does the study provide a clear description of the techniques used?  \\ 
        QAC3 &  Are the results clearly described and interpreted in the context of the objectives?  \\ 
        QAC4 &  Does the study make a contribution to the academic or industrial community?  \\ 
  \bottomrule
\end{tabular}
\end{table}

\subsection{Snowballing Search} \label{sec:snwb}
To recover relevant studies that the keyword search may have missed, we performed a snowballing search following the guidelines of Wohlin~\cite{10.1145/2601248.2601268}. Snowballing inspects an article's reference list, known as backward snowballing, and the articles that cite it, known as forward snowballing. Applied to the studies retained after quality assessment, backward and forward snowballing produced \texttt{2,002} and \texttt{4,830} articles respectively, for a total of \texttt{6,832}. We retrieved the references and the citing articles of each study from OpenAlex\footnote{\url{https://openalex.org}} and Google Scholar\footnote{\url{https://scholar.google.com}}. After deduplication against the studies already examined, we applied the same screening, inclusion and exclusion criteria, and quality assessment to these articles (Section~\ref{sec:ss}). This single iteration added \texttt{31} new studies and brought the final corpus to \texttt{122}.

\subsection{Data Extraction and Synthesis} \label{sec:dea}
Data extraction followed a predefined review protocol, available in our replication package. For each primary study, we extracted the content fields listed in Table~\ref{tab:extraction}, each linked to a research question.

We extracted and verified each study against the protocol and resolved disagreements as the protocol specifies. For the four classification fields, namely research objective, method, vulnerability class, and attack pattern, we applied open coding~\cite{799955}. We assigned codes to the relevant passages of each study and grouped codes of similar meaning into categories, whose definitions form a codebook released in our replication package. For the validation field, each value was extracted from the full text together with the passage that supports it, and values were re-checked against these passages and corrected where needed. An artifact counts as accessible when the paper declares it available and its link resolves.

We then synthesized the extracted data per research question. The coded categories became the taxonomies presented in Section~\ref{sec:rslt}, and for each research question, we report how the studies distribute across categories and discuss the dominant patterns.

\begin{table}[ht!]
\caption{Data extraction form: content fields}
\label{tab:extraction}
\centering
\begin{tabular}{p{0.26\textwidth}p{0.48\textwidth}c}
\toprule
\textbf{Field} & \textbf{Description} & \textbf{RQ} \\
\midrule
SDN component(s) & Software component(s) addressed: SDN controller, 
SDN application, interface (northbound, southbound, east-west), or data plane & RQ1 \\
Research objective & Main security objective of the study & RQ1 \\
Validation & How the study validates its contribution (items in Table~\ref{tab:research-practice}) & RQ1 \\
Method / technique & Testing or analysis technique used & RQ2 \\
Tool & Name of the tool or artifact proposed, if any & RQ2 \\
Vulnerability class & Class(es) of vulnerability addressed & RQ3 \\
Attack pattern & Attack pattern(s) described & RQ4 \\
Exploited vulnerability & Vulnerability class exploited by the attack, as stated in the study & RQ4 \\
Mitigation / defense & Defense proposed against the attack, or ``not reported'' & RQ4 \\
\bottomrule
\end{tabular}
\end{table}

%% file: Sections/4-result.tex
\subsection{Corpus Overview}\label{sec:overview}
The final corpus comprises 122 primary studies published between 2012 and 2026. Most are conference papers, accounting for 67 studies (55\%), followed by 44 journal articles (36\%) and 11 workshop papers (9\%); Table~\ref{tab:distpub} reports the full per-venue distribution. We analyze this corpus along the four dimensions set by our research questions. RQ1 examines the current trends, RQ2 the methods used to test and analyze SDN software for security defects, RQ3 a taxonomy of SDN software vulnerabilities, and RQ4 the attack patterns that exploit them, together with the software-level defenses reported against each.

\begin{table}[ht!]
\caption{Distribution of Publications Based on Conference, Journal Venues, and Workshop}
\label{tab:distpub}
\centering
\input{Table/Primary_studies}
\end{table}

\subsection{RQ1: What are the current trends in the literature on software security in SDN?} \label{sec:rq1result}
This question characterizes research trends from two angles, namely how the literature has evolved over time and which security topics it covers across the SDN software components.

\subsubsection{Temporal trends}
The corpus spans 2012 to 2026 (Figure~\ref{fig:rq1a}). Output grew steadily from 2012 to a sustained peak between 2017 and 2019, when the field produced 14 to 15 studies a year and reached its high of 15 in 2019. It then fell to a low of 4 studies in 2023 before recovering to 10, 9 and 8 in 2024, 2025 and 2026. More than half of the corpus, 70 of 122 studies, appeared in 2019 or later. The venue mix shifts in step with this growth. The early years from 2013 to 2015 contain only conference papers; journal articles first appear in 2016 and become prominent from 2019 onward, with 34 of the 44 journal papers published in that window, while workshop papers stay occasional and stop after 2022. The thematic focus matured as well, moving from the early dominance of access control, threat modeling, and verification toward vulnerability discovery through testing and program analysis, whose share of annual output grew from under a tenth before 2017 to roughly a fifth since 2020.

\begin{figure}[htbp!]
\centering
\input{Figures/rq1_temporal}
\end{figure}

\subsubsection{Security topics covered by the literature}
Research concentrates on the controller, which 72 of the 122 studies address, and the single most studied theme is access control, the subject of 35 studies that govern which applications may invoke control-plane resources. The interfaces (24 studies), applications (13), and data plane (13) receive less attention, as shown in Figure~\ref{fig:slr_rq1_topics}. We summarize each component in turn.

\noindent
\textbf{SDN Controller.} As the central software artifact of the control plane, the controller is examined through the full range of software-security activities, most often vulnerability discovery, with 15 studies, followed by access control with 14 and mitigation with 13. Access-control studies add software permission, authorization, and trust-management layers that constrain how applications invoke controller APIs~\cite{8865651}. Mitigation studies re-engineer the controller for security through non-bypassable enforcement kernels such as FortNOX~\cite{10.1145/2342441.2342466} and SE-Floodlight~\cite{Porras2015}, security-hardened network operating systems such as ROSEMARY~\cite{10.1145/2660267.2660353}, and a centralized root of trust in ANCHOR~\cite{10.1145/3301305}. Vulnerability-discovery studies apply fuzzing and static analysis to controller code and core services~\cite{10.1145/3243734.3243799, Ujcich2020, 10.1109/TNET.2022.3140824, Ollando_2026} and surface new software-flaw classes, including harmful race conditions~\cite{Xu2017451} and buffered-packet hijacking~\cite{Cao2020}. The remaining controller work is sparser. Several studies demonstrate control-plane attacks mounted from compromised applications, switches, or hosts, among them cross-app poisoning and topology poisoning~\cite{10.1145/3243734.3243759, 9152642, 10.1145/3658644.3670301, 10.14722/ndss.2015.23283}. Others assess controller security through threat modeling, or through automated penetration testing, as in DELTA~\cite{Lee2017}, or formally verify controller programs and protocols, as in VeriCon~\cite{10.1145/2594291.2594317, 10.1145/3319535.3363214}. A handful of studies run malicious-behavior detection through data provenance~\cite{DAM2025104677, Liu_2025}, and one performs forensic root-cause analysis with PICOSDN~\cite{272212}. RQ2 details the techniques behind this work, RQ3 the vulnerabilities they expose, and RQ4 the attacks and defenses.

\noindent
\textbf{Interfaces.} Interface research centers on access control of the northbound API, with 19 of the 24 studies governing which applications may invoke controller resources. These range from the early delegated-control API of PANE~\cite{10.1145/2486001.2486003} to permission-based, attribute-based, and blockchain-based authorization schemes. The five remaining studies analyze the OpenFlow protocol with STRIDE threat modeling~\cite{6733671}, harden the southbound channel in software~\cite{9942644, Hartono_2026}, formally verify an OpenFlow mechanism~\cite{8377889}, and demonstrate that a third-party application can identify IoT devices from the flow statistics the northbound API exposes to it~\cite{Serrat_2026}.

\noindent
\textbf{SDN Applications.} Application-level research comprises only 13 studies spread thinly across seven themes. Some verify application code, including a machine-checked Coq toolchain~\cite{10.1007/978-3-319-03545-1_3}. Others detect malicious applications before deployment, as INDAGO does~\cite{8526819}, model application permissions with role-based access control~\cite{8888031}, or assess application security through STRIDE threat modeling~\cite{10.1145/2684793.2684797}. The remaining studies each treat, in isolation, a single application-level vulnerability~\cite{10330440}, mitigation~\cite{9789775}, or attack~\cite{10.1007/s11036-023-02156-0}, which leaves the application layer the most fragmented and least consolidated part of the field.

\noindent
\textbf{Data Plane.} Data-plane research targets the software of programmable switches built with P4 or eBPF and of virtual switches. Six of its 13 studies are exploitation, subverting packet processing on these switches, with examples that range from probabilistic adversarial testing of stateful P4 programs in P4wn~\cite{10.1145/3445814.3446764} to code-execution attacks on cloud virtual switches~\cite{10.1145/3185467.3185468} and adversarial manipulation of eBPF stacks~\cite{10.1109/cns66487.2025.11194984}. The other seven pursue defense and assurance. This includes vulnerability discovery by fuzzing in CHIMERA~\cite{10.1109/sp61157.2025.00194}, verification of P4 and packet-processing programs in Vera~\cite{10.1145/3230543.3230548} and software-dataplane verification~\cite{10.5555/2616448.2616459}, software hardening through a covert-channel shim~\cite{8428482} and program obfuscation~\cite{10.1109/tdsc.2023.3277939}.

\subsubsection{How the studies validate their contributions}
Table~\ref{tab:research-practice} summarizes how the primary studies validate their contributions. Most studies evaluate on a controller implementation, and 98 studies execute the implementation. Nineteen studies report that the developers confirmed or fixed the reported defects, or that the defects were responsibly disclosed, and 11 of these obtained CVE identifiers. Only 3 of the 24 studies published before 2017 compare their results with an earlier technique or tool, against 14 of the 27 studies published since 2024. The first frameworks had no comparable predecessor, whereas later controller fuzzers compare with DELTA and BEADS (Section~\ref{sec:rq2result}). The artifact of 16 studies is publicly available.

\begin{table}[t]
  \centering
  \caption{How the 122 primary studies validate their contributions. Counts are out of 122 unless a denominator is given; a study may evaluate on several kinds of software.}
  \label{tab:research-practice}
  \begin{tabular}{@{}llr@{}}
    \toprule
 Facet & Value & \#studies \\
    \midrule
 Contribution & Proposes a technique or tool & 86 \\
                 & Demonstrates an attack & 20 \\
                 & Assesses existing software & 9 \\
                 & Builds a formal model & 4 \\
                 & Proposes a design without implementation & 3 \\
    \midrule
 Software used in the evaluation & Controller implementation & 92 \\
                 & Data-plane software (switch, compiler, P4 or eBPF program) & 26 \\
    \midrule
 Evaluation & Executes the implementation & 98 \\
                 & Does not execute an implementation & 24 \\
    \midrule
 Compares with an earlier technique or tool & 2012--2016 & 3 of 24 \\
                 & 2017--2019 & 5 of 43 \\
                 & 2020--2023 & 7 of 28 \\
                 & 2024--2026 & 14 of 27 \\
    \midrule
 Defects & Confirmed or fixed by the developers, or responsibly disclosed & 19 \\
                 & \quad with CVE identifiers & 11 \\
    \midrule
     Artifact accessibility & Publicly available & 16 \\
    \bottomrule
  \end{tabular}
\end{table}

\subsubsection{Why research concentrates on the controller}
Three factors explain the concentration on the controller, and the corpus supports each of them. First, controller implementations matured earlier than programmable data planes. NOX was released with OpenFlow in 2008~\cite{mckeown2008openflow, gude2008nox}, whereas the P4 language was published in 2014~\cite{bosshart2014p4} and the eBPF-based XDP data path in 2018~\cite{hoiland2018xdp}. In the corpus, the first controller study dates from 2012~\cite{10.1145/2342441.2342466}, the first study of a P4 data plane from 2018~\cite{10.1145/3230543.3230548}, and the first attacks on P4 or eBPF data planes from 2021~\cite{9464034, 10.1145/3445814.3446764}. Second, the controller is the most accessible experimental target. Even studies of applications and interfaces evaluate on a controller implementation in 25 of 37 cases. Third, tools accumulate where earlier tools exist, because each one becomes the reference against which the next is evaluated. The longest lineage lies within the controller, from DELTA and BEADS to FSF, FuzzSDN, and SeqFuzzSDN~\cite{Lee2017, 10.1007/978-3-319-66332-6_14, 10.1007/978-3-030-39303-8_4, 10.1145/3641541, Ollando_2026}. Data-plane lineages are recent and short, from Vera to P4wn~\cite{10.1145/3230543.3230548, 10.1145/3445814.3446764} and from analysis-guided attacks to the obfuscation that defeats them~\cite{10.1109/netsoft54395.2022.9844121, 10.1109/tdsc.2023.3277939}. The controller's share is therefore stable, between 50 and 65 percent of the studies in each of the four periods of the corpus.

\input{Figures/sunburst}


\begin{answerbox}
\textbf{\color{rqa}Answer to RQ1:} \emph{Research on SDN software security grew from 2012 to 2019, declined between 2020 and 2023, and has recovered since 2024, now appearing in journals as well as conferences in a pattern typical of a maturing field. Its emphasis has shifted from securing SDN software by design, through access-control architectures, threat modeling, and formal verification, toward discovering vulnerabilities in deployed implementations through testing and program analysis. The controller remains the focal artifact, addressed by 72 of the 122 studies, and access control is the most studied concern, with 35. Application-level security, data-plane defense, and runtime detection and forensics remain the least explored, and they are the clearest opportunities for future work.}
\end{answerbox}

\subsection{RQ2: What methods are used to test and analyze security vulnerabilities in SDN software components?} \label{sec:rq2result}
This question turns from \emph{what} the literature studies to \emph{how} it finds defects. We assigned each study to the technique families its authors applied, either as the main contribution or as a step of their solution or evaluation; techniques that are only discussed, manual attack demonstrations, and runtime defenses without such a step are not coded, the last two being covered by RQ4. Dynamic testing executes the software and observes failures, static program analysis inspects code without running it, formal methods prove or refute properties against a model, and learning-based detection and threat modeling respectively infer malicious behavior from observed activity and enumerate threats systematically. The use made of observed behavior separates testing from detection: approaches that learn from it to generate test inputs~\cite{10534286, 10.1145/3641541, Ollando_2026} are testing, whereas learning, provenance, or sandboxing that classifies or reconstructs it~\cite{9359040, DAM2025104677, 8109275, 8526819, Liu_2025, Sinha_2026, 272212} is detection. Under these criteria, 53 of the 122 studies apply at least one family, for 57 assignments (Table~\ref{tab:rq2-matrix}; studies per cell in Table~\ref{tab:rq2-refs}). Forty-nine studies apply a single family and four apply two, pairing static analysis~\cite{8526819, Liu_2025} or model checking~\cite{8109275} with learning-based detection, or threat modeling with fuzzing~\cite{10.1145/3427796.3427798}; no study applies three. We discuss each family in turn.

\begin{table}[t]
  \centering
  \caption{Component--method matrix (RQ2): number of studies applying each technique family to each SDN software component. A study counts once per family it applies, so assignments exceed studies for components with two-family studies.}
  \label{tab:rq2-matrix}
  \begin{adjustbox}{width=\linewidth,center}
  \begin{tabular}{@{}lrrrrrr@{}}
    \toprule
    Component & Dynamic testing & Static analysis & Formal methods & Learning \& threat modeling & \#studies & \#assignments \\
    \midrule
    Controller   & 15 &  5 &  6 &  6 & 31 & 32 \\
    Applications &  0 &  2 &  3 &  5 &  8 & 10 \\
    Interfaces   &  0 &  1 &  3 &  1 &  5 &  5 \\
    Data plane   &  5 &  2 &  2 &  1 &  9 & 10 \\
    \midrule
    Total        & 20 & 10 & 14 & 13 & 53 & 57 \\
    \bottomrule
  \end{tabular}
  \end{adjustbox}
\end{table}

\begin{table}[t]
  \centering
  \caption{Studies per component and technique family (RQ2).}
  \label{tab:rq2-refs}
  \small
  \begin{tabular}{@{}llp{8.5cm}@{}}
    \toprule
    Component & Family & Studies \\
    \midrule
    Controller   & Dynamic testing  & \cite{8718586, 10.1145/3243734.3243799, 10.1007/978-3-319-66332-6_14, 10.1007/978-3-030-39303-8_4, 10534286, Kim20234463, LEE2020101720, 10.1109/TNET.2022.3140824, Lee2017, 10988999, 10.1145/3641541, Ollando_2026, 8023155, 8725649, Xu2017451} \\
                 & Static analysis  & \cite{Liu_2025, 10.1145/3243734.3243759, Ujcich2020, 10.1155/2017/8740217, 9152642} \\
                 & Formal methods   & \cite{10.1145/2594291.2594317, 9633403, 10.1145/3301305, 10.1145/3319535.3363214, 10.1007/s11036-018-1141-9, 10494990} \\
                 & Learning \& threat modeling & \cite{7751150, DAM2025104677, 10.1007/978-3-319-71761-6_6, 10.1145/2684793.2684796, Liu_2025, 272212} \\
    \midrule
    Applications & Static analysis  & \cite{10.1145/2876019.2876026, 8526819} \\
                 & Formal methods   & \cite{8109275, 10.1007/978-3-319-03545-1_3, YAO2025111259} \\
                 & Learning \& threat modeling & \cite{9359040, 8109275, 8526819, Sinha_2026, 10.1145/2684793.2684797} \\
    \midrule
    Interfaces   & Static analysis  & \cite{HU2021108} \\
                 & Formal methods   & \cite{10.1007/s12243-025-01133-w, LENG201968, 8377889} \\
                 & Learning \& threat modeling & \cite{6733671} \\
    \midrule
    Data plane   & Dynamic testing  & \cite{10.1145/3427796.3427798, 10.1145/3445814.3446764, 10.1109/sp61157.2025.00194, 10.1145/3140649.3140651, 10.1145/3185467.3185468} \\
                 & Static analysis  & \cite{10.1109/netsoft54395.2022.9844121, 10.1109/tdsc.2023.3277939} \\
                 & Formal methods   & \cite{10.5555/2616448.2616459, 10.1145/3230543.3230548} \\
                 & Learning \& threat modeling & \cite{10.1145/3427796.3427798} \\
    \bottomrule
  \end{tabular}
\end{table}

\subsubsection{Dynamic testing}
Dynamic testing is the most active technique in the corpus, and fuzzing dominates it. Controller fuzzers mutate the inputs that drive the control plane, most often OpenFlow messages and control-plane events, and observe the controller for crashes, hangs, and policy violations. BEADS generates close to nineteen thousand attack scenarios from malformed or misdelivered OpenFlow and ARP messages and uncovers 831 unique bugs across Ryu, POX, Floodlight, and ONOS~\cite{10.1007/978-3-319-66332-6_14}, while FSF adds code-coverage feedback to the same setting and surfaces bugs that black-box fuzzing leaves untouched~\cite{10.1007/978-3-030-39303-8_4}. More recent fuzzers target deeper behaviors: SPIDER fuzzes for stateful performance defects in ONOS~\cite{10988999}, Ambusher applies protocol state fuzzing to the east-west channel of distributed controllers~\cite{10534286}, and AIM-SDN stresses the controller datastore to expose information-mismanagement flaws~\cite{10.1145/3243734.3243799}. Fuzzing extends to the data plane as well, where P4Fuzz fuzzes P4 compilers~\cite{10.1145/3427796.3427798} and P4wn profiles stateful P4 programs probabilistically to drive adversarial tests~\cite{10.1145/3445814.3446764}. Beyond fuzzing, ATTAIN injects faults into the control channel to evaluate controller resilience~\cite{8023155}, CONGUARD detects harmful race conditions between concurrent control-plane event handlers~\cite{Xu2017451}, DELTA orchestrates penetration tests that replay known attacks and randomly explore for new ones~\cite{Lee2017}, and a recent line learns failure-inducing input models to test controllers more efficiently~\cite{10.1145/3641541} and extends them to stateful behavior by inferring extended finite-state machines that guide the fuzzing of control-message sequences on ONOS and Ryu~\cite{Ollando_2026}. The common thread is that execution-based techniques excel at the controller and the programmable data plane, where concrete inputs are abundant, and crashes are unambiguous oracles.

\subsubsection{Static program analysis}
Static analysis reasons about SDN software without executing it, which lets it scale across whole codebases and reach defects that testing rarely triggers. At the application level, SHIELD builds the control-flow graph of an SDN application and extracts its critical flows to flag malicious behaviors before the app is loaded onto the controller~\cite{10.1145/2876019.2876026}. Other results in the corpus come from analyzing how control-plane events and state flow between components. Cross-plane analysis builds an event-flow graph of the controller and discovers event-based vulnerabilities that no single-component tool would see~\cite{Ujcich2020}, and information-flow analysis exposes cross-app poisoning, where one application influences another through shared datastore state~\cite{10.1145/3243734.3243759}. Static techniques also drive attack discovery on the data plane: static-analysis-guided attacks identify exploitable paths in programmable data planes~\cite{10.1109/netsoft54395.2022.9844121}, and obfuscation has been proposed precisely to blunt such analysis when an adversary wields it~\cite{10.1109/tdsc.2023.3277939}. Static analysis is therefore the technique of choice for the interaction defects that this review identifies as a main hazard of SDN software.

\subsubsection{Formal methods}
Formal methods give the strongest guarantees in the corpus, at the cost of a model that abstracts the real implementation. Controller verification is the largest subgroup. VeriCon proves safety invariants of controller programs over all admissible topologies and event sequences~\cite{10.1145/2594291.2594317}, Proof-Carrying Network Code attaches machine-checkable proofs to network programs so a controller can reject code that violates policy~\cite{10.1145/3319535.3363214}, and formal verification has been used to discover attacks in controllers automatically rather than merely to confirm their absence~\cite{10494990}; CSP models verify core modules of Floodlight~\cite{10.1007/s11036-018-1141-9}, and ProVerif checks an SDN access-control protocol for authentication flaws~\cite{9633403}. Verification also reaches the other components. On the application layer, a machine-checked Coq toolchain verifies NetCore programs end to end~\cite{10.1007/978-3-319-03545-1_3}, model checking tests security properties of protocols under development~\cite{YAO2025111259}, and software model checking with a customized Java PathFinder, combined with analytics over execution data, verifies SDN application code~\cite{8109275}. On the interfaces, a CSP model captures and verifies the OpenFlow scheduled-bundle mechanism~\cite{8377889}. On the data plane, Vera symbolically executes P4 programs to debug them~\cite{10.1145/3230543.3230548} and software-dataplane verification proves properties of packet-processing code~\cite{10.5555/2616448.2616459}.

\subsubsection{Learning-based detection and threat modeling}
The final family does not search code for defects but watches software behavior or reasons about threats systematically. Learning-based detectors flag malicious SDN applications: INDAGO classifies them before deployment from static and behavioral features~\cite{8526819}, and SandboxNet detects them online from their runtime activity~\cite{9359040}. Provenance-based detection records the causal history of control-plane actions to expose malicious behavior after the fact~\cite{DAM2025104677}, ProvGuard builds the provenance graph of the controller's own activities, instrumented through static analysis of its code, to detect control-policy manipulation originating from the data plane~\cite{Liu_2025}, and PICOSDN turns the same provenance data toward forensic root-cause analysis of attacks~\cite{272212}. Alongside these, several studies apply structured threat modeling, most often STRIDE, to enumerate threats against controllers, applications, and the OpenFlow protocol~\cite{10.1007/978-3-319-71761-6_6, 10.1145/2684793.2684797, 6733671}. Threat modeling is cheap and component-agnostic, which explains its breadth across the corpus, but it surfaces hypothesized threats rather than confirmed defects, and only one study in the corpus pairs it with another family~\cite{10.1145/3427796.3427798}.

\subsubsection{Evolution of the technique families}
The four families appeared at different times and followed different routes. Formal methods appeared in 2013--2014, with proofs over network programs, controller programs, and a software data plane~\cite{10.1007/978-3-319-03545-1_3, 10.1145/2594291.2594317, 10.5555/2616448.2616459}. From 2018 they extended to controller modules, OpenFlow mechanisms, P4 programs, and SDN security protocols~\cite{8377889, 10.1007/s11036-018-1141-9, 10.1145/3230543.3230548, 10.1145/3301305, 9633403}, and from 2024 they were used to discover attacks~\cite{10494990, YAO2025111259}. Threat modeling also appeared early, with STRIDE analyses from 2013 to 2018~\cite{6733671, 10.1145/2684793.2684796, 10.1145/2684793.2684797, 7751150, 10.1007/978-3-319-71761-6_6}; after 2018 it appeared only as a step of another technique~\cite{10.1145/3427796.3427798}. Static analysis started in 2016 with single applications~\cite{10.1145/2876019.2876026}, turned to interactions between applications, events, and data dependencies in 2018--2020~\cite{10.1145/3243734.3243759, Ujcich2020, 9152642}, and reached programmable data planes in 2022~\cite{10.1109/netsoft54395.2022.9844121}. Dynamic testing started in 2017 with black-box frameworks~\cite{Lee2017, 10.1007/978-3-319-66332-6_14, 8023155}. From 2020, fuzzers added feedback from the controller under test: code coverage~\cite{10.1007/978-3-030-39303-8_4}, then intent-state transitions, learned failure models, inferred protocol state machines, and message sequences~\cite{Kim20234463, 10.1145/3641541, 10534286, Ollando_2026}. In 2025, fuzzing spanned both planes, because many P4 defects arise from their interplay~\cite{10.1109/sp61157.2025.00194}. Behavior-based detection started in 2017--2018~\cite{8109275, 8526819} and moved to data provenance from 2021~\cite{272212, DAM2025104677, Liu_2025}; ProvGuard adopted it because detection that watches only the data plane misses manipulations of the controller's state~\cite{Liu_2025}. Across families, P4 data planes entered late, in 2018 for formal methods, 2021 for dynamic testing, and 2022 for static analysis~\cite{10.1145/3230543.3230548, 10.1145/3427796.3427798, 10.1109/netsoft54395.2022.9844121}, consistent with the later maturation of programmable data planes (Section~\ref{sec:rq1result}). Techniques that span components appeared late within their families, with cross-plane event analysis in 2020 and cross-plane fuzzing in 2025~\cite{Ujcich2020, 10.1109/sp61157.2025.00194}.

\begin{answerbox}
\textbf{\color{rqa}Answer to RQ2:} \emph{The literature finds SDN software defects with four families of techniques. Dynamic testing, led by OpenFlow and event fuzzing, is the most active and concentrates on the controller and the programmable data plane, where concrete inputs and crash oracles are plentiful. Static program analysis is the method of choice for the cross-plane and cross-app interaction defects that single-component tools miss. Formal methods give the strongest guarantees and span every component, from Coq-verified applications to symbolically executed P4 programs, but depend on faithful models. Learning-based detection and threat modeling complete the picture, the former watching runtime behavior and the latter enumerating threats systematically, though threat modeling alone yields hypotheses rather than confirmed defects. Techniques are rarely combined: four of the 53 coded studies apply two families, and none applies more.}
\end{answerbox}

\subsection{RQ3: How can SDN software vulnerabilities be classified?} \label{sec:rq3result}
Having established what the literature studies (RQ1) and how it finds defects (RQ2), we now organize the defects themselves. We coded the vulnerability class of every study that reports or addresses a concrete weakness and grouped the codes by the SDN software component in which the weakness resides. The result is the taxonomy in Figure~\ref{fig:txnmvul}, which mirrors the software architecture: vulnerabilities of the controller, of the applications, of the interfaces, and of the data plane. The taxonomy lets a practitioner locate the weaknesses that pertain to a given component, and a researcher see which classes are well studied and which are not. Two cross-cutting observations frame what follows. First, a large share of the reported weaknesses are not memory-safety bugs but design and logic flaws, namely, missing authorization, absent isolation, and unsafe component interaction. Second, because control is centralized, a weakness anywhere in this taxonomy can have network-wide consequences, which RQ4 makes concrete.

\begin{figure}[!ht]
\begin{adjustbox}{width=\linewidth, center}
    \centering
    \begin{forest}
        forked edges, folder indent=1cm,
        where={level()==0}{}{folder, grow'=east, rounded corners},
        where={level()>0}{l sep+=1cm}{
        draw=rq, thick, fill=rq, rounded corners,
        },
        for tree={
            fork sep=3mm, thick, edge=thick,
            if n children=0{if n=1{yshift=-5mm}{}, for parent={s sep=1mm}}{draw, minimum height=6ex, minimum width=4cm}
        },
        calign=child edge, calign child=2
        [SDN software vulnerabilities
            [Controller
                [Software bugs and races]
                [Unsafe component interaction]
                [Missing authorization]
                [No resource control]
                [Single point of failure]
            ]
            [Applications
                [No application authentication]
                [Inadequate sandboxing]
                [Over-privileged applications]
                [Insecure development]
                [Application interference]
            ]
            [Interfaces
                [Missing authorization]
                [Weak authentication]
                [Excessive privileges]
                [Insufficient encryption]
                [Insecure API design]
            ]
            [Data plane
                [Programmable-switch bugs]
                [Unsafe packet parsing]
                [Stateful-pipeline misuse]
                [Missing integrity check of loaded programs]
            ]
        ]
    \end{forest}
\end{adjustbox}
\caption{Taxonomy of SDN software vulnerabilities, organized by software component.}
\label{fig:txnmvul}
\end{figure}

\subsubsection{Controller vulnerabilities}
The controller is the most vulnerability-rich component because it is the trusted code at which every plane converges. The most direct weaknesses are ordinary software defects, since the controller integrates code from many developers, and any of it may contain bugs an attacker can reach~\cite{Kim20234463, Ujcich2020, Ollando_2026}. Concurrency makes this worse: the control plane processes events on multiple threads, and harmful race conditions between event handlers corrupt shared state in ways that single-threaded reasoning never anticipates~\cite{Xu2017451}. A second, subtler class is unsafe component interaction, where no single module is wrong but their composition is, producing event-based and data-dependency vulnerabilities that conventional tools miss~\cite{Ujcich2020, 9152642}. The controller also inherits operating-system-style weaknesses that legacy controllers omitted, namely, missing authorization that lets applications perform privileged actions unchecked~\cite{10.1109/TNET.2017.2748159, 10.1145/2660267.2660353} and the absence of resource control that lets a single application exhaust controller CPU or memory~\cite{10.1145/2660267.2660353}. Underlying all of these is the architectural fact that the logically centralized controller is a single point of failure, so any of the preceding defects can escalate from a local fault to a network-wide outage~\cite{10.1109/TNET.2017.2748159, 7943369}.

\subsubsection{Application vulnerabilities}
SDN applications are network programs that run with control-plane privilege, and their weaknesses stem from how loosely that privilege is governed. Most controllers in use provide no mechanism to authenticate the developer or provenance of an application, so a malicious program can install itself in the guise of a legitimate one~\cite{ALIYU2020107421}. Once installed, applications are too rarely isolated: inadequate sandboxing lets a fault or attack in one application propagate to others or to the controller itself~\cite{10.1145/2660267.2660353}. The problem compounds when applications are over-privileged, holding access to sensitive resources far beyond their function, which turns any compromise into a privilege-escalation foothold~\cite{8888031}. Applications written without secure-coding discipline introduce the familiar input-validation and dependency flaws of ordinary software, now amplified by control-plane reach~\cite{10837050}. Finally, applications interfere with one another: when they form service chains or operate on shared state, a malicious or buggy application can intercept messages, trigger processing loops, or override the decisions of its neighbors, a class of interference that dedicated analyses have had to be built to detect~\cite{10.1145/3243734.3243759}.

\subsubsection{Interface vulnerabilities}
The interfaces carry every cross-plane interaction, yet the northbound and east-west interfaces remain unstandardized, which leaves their security to each implementation. The recurring weakness is missing authorization: the northbound API exposes operations with network-wide impact but frequently fails to check whether the caller is permitted to invoke them~\cite{10.1145/2660267.2660353, 10.1145/2876019.2876024}. Weak authentication follows from the OpenFlow specification's lack of a standard authentication mechanism, which can let unauthorized entities reach the control plane~\cite{10.1109/TNET.2017.2748159}; on the east-west interface, it takes the form of a one-time authentication of peer controllers that is never re-evaluated~\cite{Shahzad_2026}. Where access control does exist, it is often too coarse, granting excessive privileges that an application can abuse to reach resources beyond its remit~\cite{7579735}. The channels themselves are sometimes inadequately encrypted, exposing flow rules and configuration to eavesdropping and manipulation in transit~\cite{10.1145/2660267.2660353}. Underlying these is insecure API design, where poorly specified interfaces admit injection, buffer, and input-validation flaws of the kind well known from public-facing software~\cite{8725649}.

\subsubsection{Data-plane vulnerabilities}
The data plane is the component that the literature addressed last and among the least studied, yet it is a software attack surface. Programmable and virtual switches are themselves programs, so P4 targets, eBPF stacks, and software switches such as Open vSwitch can contain exploitable defects. Table~\ref{tab:dataplane-profiles} compares the weaknesses that the corpus reports for each of them.

\begin{table}[t]
  \centering
  \small
  \caption{Data-plane software and the weaknesses reported for it in the corpus. Two studies target P4 programs compiled to eBPF and are counted in both columns; the two remaining data-plane studies verify a Click software data plane~\cite{10.5555/2616448.2616459} and resolve flow-rule conflicts~\cite{8428482}.}
  \label{tab:dataplane-profiles}
  \begin{tabular}{@{}l>{\raggedright\arraybackslash}p{3.2cm}>{\raggedright\arraybackslash}p{4.4cm}>{\raggedright\arraybackslash}p{3.6cm}@{}}
    \toprule
     & Open vSwitch & P4 programs and targets & eBPF stacks \\
    \midrule
    Studies & 2~\cite{10.1145/3140649.3140651, 10.1145/3185467.3185468} & 8~\cite{10.1145/3230543.3230548, 9464034, 10.1145/3427796.3427798, 10.1145/3445814.3446764, 10.1109/netsoft54395.2022.9844121, 10.1109/tdsc.2023.3277939, 10.1109/sp61157.2025.00194, Su_2026} & 3~\cite{10.1109/netsoft54395.2022.9844121, 10.1109/tdsc.2023.3277939, 10.1109/cns66487.2025.11194984} \\
    Weaknesses & Unsafe packet parsing & Unsafe packet parsing; programmable-switch bugs; stateful-pipeline misuse; cross-plane defects & Missing integrity check of loaded programs \\
    Impact & Code execution on the host, then on the controller; CVE-2016-2074, CVE-2016-10377 & Firewall bypass; forwarding changes; denial of service; two CVEs in ONOS & Denial of service; traffic manipulation \\
    Defenses & Upstream fixes & Verification, compiler fuzzing, obfuscation, consistency validation & Obfuscation; proposed countermeasures \\
    \bottomrule
  \end{tabular}
\end{table}

Whether these weaknesses form a different class from controller defects depends on the software. The Open vSwitch defects are memory-safety bugs of ordinary C code. What differs is their reach, because the switch runs as root alongside the hypervisor: one crafted packet from a virtual machine compromises the host and then the controller~\cite{10.1145/3140649.3140651, 10.1145/3185467.3185468}. P4 and eBPF programs fail differently. They process every packet at line rate with state confined to registers, tables, or maps, the pipeline structure on which the verification of software data planes relies~\cite{10.5555/2616448.2616459}. In the corpus, their defects are parser errors, invalid header accesses, state that traffic can steer, and compiler defects, rather than logic errors in event handling~\cite{10.1145/3230543.3230548, 10.1145/3427796.3427798, 10.1145/3445814.3446764}. The attacks of 2021--2025 add a fourth class, missing integrity check of loaded programs: nothing prevents or detects an attacker present on the device from editing table entries or the compiled program, and probe-based verification cannot scale beyond probabilistic detection of such edits~\cite{9464034, 10.1109/netsoft54395.2022.9844121, 10.1109/cns66487.2025.11194984}. A further group of defects belongs to neither plane. All seven defects found by Chimera require control-plane inputs, and two require inputs from both planes~\cite{10.1109/sp61157.2025.00194}; SCT-V addresses inconsistencies between the two planes rather than defects of either~\cite{Su_2026}. Unsafe packet parsing is illustrated by the unified parser of Open vSwitch: a packet whose MPLS label stack exceeds the parser's threshold overflows a stack buffer and gives the attacker control of the physical machine~\cite{10.1145/3140649.3140651}. Stateful-pipeline misuse is illustrated by a port-knocking firewall written in P4, whose register is indexed by a per-address identifier. An attacker who can edit table entries reassigns to its own address the identifier of an authorized address and inherits its access rights~\cite{9464034}; this case also involves a missing integrity check. P4wn reaches harmful states from traffic alone, with inputs that cause hash collisions and concentrate flows on one port~\cite{10.1145/3445814.3446764}. These weaknesses matter because data-plane code executes attacker-influenced input at line rate, with the same potential to compromise forwarding that controller defects have to compromise control.

\begin{answerbox}
\textbf{\color{rqa}Answer to RQ3:} \emph{The vulnerabilities reported across the corpus organize by software component. Controller weaknesses are dominated by software bugs, control-plane races, and unsafe component interaction, layered over missing authorization, absent resource control, and the single point of failure. Application weaknesses trace to ungoverned privilege, namely, missing application authentication, weak sandboxing, over-privilege, insecure development, and inter-application interference. Interface weaknesses center on missing authorization and weak authentication, compounded by excessive privilege, insufficient encryption, and insecure API design. Data-plane weaknesses, though among the least studied, are real software flaws in programmable and virtual switches, especially unsafe packet parsing, stateful-pipeline misuse, and missing integrity checks of loaded programs. Most of these are design and logic flaws rather than memory-safety bugs, and because control is centralized, any of them can escalate to network-wide impact.}
\end{answerbox}

\subsection{RQ4: What attack patterns target SDN software vulnerabilities, how do they exploit these weaknesses, and what defenses mitigate them?} \label{sec:rq4results}

RQ3 cataloged the weaknesses; this question asks how attackers turn them into compromises and how the literature defends against them. For every study that describes an attack, we coded the attack pattern, the vulnerability class it exploits, and the software-level defense it reports, leaving the defense field empty when none is given. We present the patterns grouped by the component they target, and Table~\ref{tab:masv} maps each pattern to the RQ3 vulnerabilities it exploits and the breadth of its impact. Two properties recur. Most patterns chain a weak interface to a privileged action, and most of the demonstrated impact is network-wide rather than local, which follows directly from the centralization that RQ1 and RQ3 emphasize.

\subsubsection{Application-originated attacks}
The richest class of attacks comes from malicious or compromised applications, because an application already holds control-plane privilege and needs only to abuse it. The simplest pattern is illegal function calling, in which an application invokes a built-in controller function it should not reach; demonstrations show that calling \texttt{System.exit()} through an unprotected northbound interface shuts down Java controllers such as Floodlight, ONOS, and OpenDaylight, and the same channel can terminate a benign application such as a firewall~\cite{Lee2017, 10.1145/2660267.2660353, 10.1145/2876019.2876024}. A subtler pattern is cross-app poisoning, where a malicious application seeds spoofed events that corrupt shared control-plane state, so that an honest forwarding application later installs the attacker's rules; flow-level access control misses this because it does not track information flow between applications~\cite{10.1145/3243734.3243759, Alsalamh2024}. Applications with global visibility mount unauthorized-access attacks, clearing packet counters to hide a denial of service or rewriting topology and flow-rule state in the controller's datastore~\cite{7943369, 10.1145/2660267.2660353}. Where applications form service chains, a single one can jam the chain by entering an infinite loop and blocking every downstream handler~\cite{10.1145/2660267.2660353}, and an over-privileged application can abuse its authority to issue system commands that disconnect APIs or other applications~\cite{10.1007/978-3-319-26362-5_16}. The defenses map onto RQ3's missing-authorization and weak-isolation classes. Permission and authorization layers constrain which APIs an application may call, as in the read, notification, write, and system categories of SE-Floodlight and ControllerSEPA~\cite{Porras2015, 6980437, 7943369}; role-based access control narrows privilege further~\cite{8888031}; SDNShield reconciles configurable per-application permissions for app markets~\cite{7579735}; and process- or Java-sandboxing isolates applications so that a fault cannot escape its host~\cite{10.1145/2660267.2660353}. Detectors complement these by flagging malicious applications before deployment~\cite{8526819} and hidden network manipulations at runtime~\cite{10.1145/3229616.3229620}.

\subsubsection{Controller-targeted attacks}
A second class targets the controller directly, aiming at availability and integrity. Controller saturation exhausts processing capacity through flooding or by triggering an expensive control-plane path, degrading or disabling the controller and, with it, the network~\cite{10.1145/3243734.3243799, 10.1145/2660267.2660353}. The reported defenses combine rate limiting~\cite{10.1145/3243734.3243799} and northbound access control that caps how aggressively an application may drive the control plane~\cite{7997249, 10.1007/s11280-022-01130-2, 10.1145/3134600.3134603}. A more insidious pattern is network-operating-system misuse, in which a malicious application installs an SDN rootkit that rewrites the controller's internal data structures, hides injected or deleted flow rules, and opens a covert OpenFlow channel for command and control~\cite{10.1007/978-3-319-26362-5_16}. SDN-Guard counters this class by protecting the controller's internal state from rootkit manipulation~\cite{8169856}. Code injection rounds out the class, where weak OpenFlow authentication, unsecured REST endpoints, or a vulnerable application let an attacker insert instructions that the controller then executes; input validation, secure coding, and patching are the reported mitigations~\cite{10837050}. In multi-domain deployments, controller hijacking adds a pattern in which a compromised or fake controller, once admitted by its peers, injects flow rules and false state across domains; the reported defense authenticates controllers at admission and re-admission and continuously scores their behavior, revoking the keys of a controller whose flow activity deviates from its profile~\cite{Shahzad_2026}. These attacks exploit the software-bug, missing-resource-control, and single-point-of-failure classes of RQ3, together with the weak authentication of peer controllers, and their impact is consistently network-wide.

\subsubsection{Interface-mediated attacks}
The third class abuses the channels between planes. Because the controller dispatches events to applications over the northbound interface, an adversary positioned on that channel can mount a man-in-the-middle attack, intercepting an event before delivery to alter its payload or suppress it entirely~\cite{Lee2017, 10.1145/2660267.2660353}. API abuse exploits coarse or absent authorization on the northbound API: with update and add permissions, an application redirects traffic through attacker-controlled paths, and with insufficient checks, it can terminate the controller~\cite{7997249, HU2021108, 8528480, 10.1145/2660267.2660353}. An inference attack requires no such privilege: an application that is authorized to read flow statistics through the northbound API can identify the connected devices and infer user activity from those statistics alone, without modifying any network state~\cite{10330440, Serrat_2026}. The defenses are squarely the access-control work that dominates the interface literature. Controller DAC adds dynamic access control at sub-percent performance overhead~\cite{7997249}, SEAPP governs REST-API access through permission detection and a registration authority~\cite{HU2021108}, blockchain-based schemes provide tamper-evident cross-domain authorization~\cite{DUY2022103080, 8528480, 9162669}, channel encryption, such as HTTPS, closes the man-in-the-middle vector~\cite{Banse2015834}, and a middleware that perturbs the statistics returned to untrusted applications reduces the inference attack~\cite{Serrat_2026}. These patterns exploit the missing-authorization, weak-authentication, excessive-privilege, and insufficient-encryption classes of RQ3.

\subsubsection{Data-plane-originated attacks}
The final class originates in the data plane. A compromised programmable or virtual switch can run attacker code: taking control of cloud systems through the data plane and the vAMP variant through the unified packet parser, both of which turn switch software into a foothold for compromising the wider system~\cite{10.1145/3185467.3185468, 10.1145/3140649.3140651}. Adversarial exploitation of P4 data planes and static-analysis-guided attacks steer programmable pipelines into harmful states~\cite{9464034, 10.1109/netsoft54395.2022.9844121}, while misreporting attacks have a compromised switch lie about traffic load, or silently fail to enforce the rules it received, to bend controller decisions in the attacker's favor~\cite{10.1007/s11036-023-02156-0, Su_2026}. Topology poisoning uses the data plane as a channel to the controller: a host or switch relays LLDP packets, migrates a host, or spoofs identifiers, and the controller's topology services accept these messages without validation and install the attacker's rules; the pattern was found in the host-tracking and link-discovery services of eight controllers~\cite{10.14722/ndss.2015.23283} and through an unprotected internal testing API~\cite{10229058}. The defenses here are still thin, comprising program obfuscation that raises the cost of analysis-guided attacks~\cite{10.1109/tdsc.2023.3277939}, a shim that catches policy violations in the data plane~\cite{8428482}, cross-app-poisoning protection extended to virtual switches~\cite{9789775}, validation triggered on forwarding-state changes that checks that switches enforce the installed rules~\cite{Su_2026}, and, against topology poisoning, plausibility checks added to the controller's topology services~\cite{10.14722/ndss.2015.23283} and anomaly detection on the provenance graph of the controller's own activities~\cite{Liu_2025}. The scarcity of mature defenses for this class is one of the clearest gaps the review identifies.

\begin{table}[ht!]
\caption{Attack patterns mapped to the exploited RQ3 vulnerability classes and to the impact scope. \textcolor{rq}{\ding{51}}~marks an exploited vulnerability; \textcolor{rd}{\ding{55}}~marks an affected component.}
\label{tab:masv}
\centering
\begin{adjustbox}{width=\linewidth,center}
\begin{tabular}{l *{7}{c} | *{4}{c}}
\toprule
\textbf{Attack pattern} & \multicolumn{7}{c}{\textbf{Exploited vulnerability}} & \multicolumn{4}{c}{\textbf{Impact}} \\
\cmidrule(lr){2-8}\cmidrule(lr){9-12}
&
\rotatebox{90}{Weak authentication\;} &
\rotatebox{90}{Missing authorization\;} &
\rotatebox{90}{Inadequate sandboxing\;} &
\rotatebox{90}{Over-privilege\;} &
\rotatebox{90}{Software bug / race\;} &
\rotatebox{90}{No resource control\;} &
\rotatebox{90}{Insufficient encryption\;} &
\rotatebox{90}{Applications\;} &
\rotatebox{90}{Interfaces\;} &
\rotatebox{90}{Controller\;} &
\rotatebox{90}{Data plane\;} \\
\midrule
Illegal function calling      & \cmk & \cmk & \cmk & \cmk &      &      &      & \xmk &      & \xmk &      \\
Cross-app poisoning           &      & \cmk &      &      & \cmk &      &      & \xmk &      & \xmk &      \\
Unauthorized access           & \cmk & \cmk &      &      &      &      &      &      &      & \xmk &      \\
Service-chain jamming         &      &      & \cmk &      &      & \cmk &      & \xmk &      & \xmk &      \\
Abuse of privilege            &      & \cmk &      & \cmk &      &      &      & \xmk & \xmk & \xmk &      \\
Controller saturation         &      & \cmk &      &      & \cmk & \cmk &      &      & \xmk & \xmk &      \\
NOS misuse / rootkit          &      &      &      &      & \cmk &      &      &      &      & \xmk &      \\
Code injection                & \cmk &      &      &      & \cmk &      &      &      & \xmk & \xmk &      \\
Controller hijacking          & \cmk &      &      &      &      &      &      &      & \xmk & \xmk &      \\
Man-in-the-middle             &      &      &      &      &      &      & \cmk &      & \xmk & \xmk &      \\
API abuse                     & \cmk & \cmk &      &      &      &      &      &      & \xmk & \xmk &      \\
Inference attack              &      &      &      & \cmk &      &      &      &      & \xmk &      &      \\
Data-plane code execution     &      &      &      &      & \cmk &      &      &      &      &      & \xmk \\
Load misreporting             &      & \cmk &      &      & \cmk &      &      &      &      & \xmk & \xmk \\
Topology poisoning            & \cmk &      &      &      & \cmk &      &      &      &      & \xmk & \xmk \\
\bottomrule
\end{tabular}
\end{adjustbox}
\end{table}


\begin{answerbox}
\textbf{\color{rqa}Answer to RQ4:} \emph{The attacks reported in the corpus fall into four groups by origin. Application-originated attacks, including illegal function calling, cross-app poisoning, unauthorized access, service-chain jamming, and privilege abuse, are the most numerous and exploit missing authorization and weak isolation. Controller-targeted attacks, namely saturation, rootkit-based operating-system misuse, code injection, and controller hijacking, exploit software bugs, absent resource control, the single point of failure, and the weak authentication of peer controllers. Interface-mediated attacks, principally man-in-the-middle, API abuse, and inference from exposed statistics, exploit weak authentication, missing authorization, excessive privilege, and insufficient encryption. Data-plane-originated attacks run code on compromised switches, misreport state to mislead the controller, and poison its view of the topology. Defenses cluster on the application and interface side, where permission models, role-based access control, sandboxing, and channel encryption are mature, but defenses for data-plane attacks remain scarce, marking the clearest gap for future work.}
\end{answerbox}

%% file: Table/Primary_studies.tex
\begin{adjustbox}{width=\linewidth,center}
\begin{tabular}{lcclcclcc}
\toprule
\textbf{Conference Venue} & \textbf{\# Studies} & \textbf{References} & \textbf{Journal Venue} & \textbf{\# Studies} & \textbf{References} & \textbf{Workshop} & \textbf{\# Studies} & \textbf{References}\\
\midrule
NDSS & 6 & \cite{Ujcich2020, Lee2017, 10.14722/ndss.2015.23283, Porras2015, Cao2020, Liu_2025} & Computer Networks & 5 & \cite{YAO2025111259, LENG201968, ALIYU2020107421, SOUSA2024110130, KANG20191} & SDN-NFVSec & 3 & \cite{10.1145/3309194.3309195, 10.1145/2876019.2876026, 10.1145/2876019.2876024} \\
CCS & 5 & \cite{10.1145/3243734.3243799, 10.1145/3243734.3243759, 10.1145/3319535.3363214, 10.1145/2660267.2660353, 10.1145/3658644.3670301} & Computers \& Security & 3 & \cite{DAM2025104677, LEE2020101720, PALADI2019155} & CCSW & 2 & \cite{Habib202223, 10.1145/3140649.3140651} \\
ICC & 3 & \cite{7997245, 7997249, 8422405} & IEEE/ACM Transactions on Networking & 3 & \cite{10.1109/TNET.2022.3140824, 8428482, 10.1109/TNET.2017.2748159} & NFV-SDN & 2 & \cite{8725649, 8169856} \\
ICNP & 3 & \cite{8526819, 6733671, 6980437} & IEEE Transactions on Dependable and Secure Computing & 2 & \cite{7976378, 10.1109/tdsc.2023.3277939} & HotSDN & 1 & \cite{10.1145/2342441.2342466} \\
USENIX Security & 3 & \cite{Xu2017451, 272212, Kim20234463} & IEEE Transactions on Information Forensics and Security & 2 & \cite{10534286, 8865651} & INFOCOM-Wksps & 1 & \cite{9162669} \\
DSN & 2 & \cite{8023155, 7579735} & IEEE Access & 2 & \cite{8343872, 8718586} & ISSREW & 1 & \cite{8109275} \\
NetSoft & 2 & \cite{7258233, 10.1109/netsoft54395.2022.9844121} & Mobile Networks and Applications & 2 & \cite{10.1007/s11036-023-02156-0, 10.1007/s11036-018-1141-9} & SecSoN & 1 & \cite{10.1145/3229616.3229620} \\
RAID & 2 & \cite{10.1007/978-3-319-66332-6_14, 10.1007/978-3-319-26362-5_16} & Security and Communication Networks & 2 & \cite{10.1155/2018/9178425, 10.1002/sec.1369} &  & &  \\
S\&P & 2 & \cite{9152642, 10.1109/sp61157.2025.00194} & ACM Transactions on Software Engineering and Methodology & 2 & \cite{10.1145/3641541, Ollando_2026} &  & &  \\
SIGCOMM & 2 & \cite{10.1145/2486001.2486003, 10.1145/3230543.3230548} & Computer Communications & 2 & \cite{ELZOGHBI2025108169, Su_2026} &  & &  \\
AINTEC & 2 & \cite{10.1145/2684793.2684797, 10.1145/2684793.2684796} &  ACM Transactions on Privacy and Security & 1 & \cite{10.1145/3301305} &  & &  \\
ACSAC & 1 & \cite{10.1145/3134600.3134603} & Annals of Telecommunications & 1 & \cite{10.1007/s12243-025-01133-w} &  & &  \\
ADHIP & 1 & \cite{10.1007/978-3-030-19086-6_60} & Concurrency and Computation: Practice and Experience & 1 & \cite{Alsalamh2024} &  & &  \\
ASPLOS & 1 & \cite{10.1145/3445814.3446764} & IEEE Internet of Things Journal & 1 & \cite{9452113} &  & &  \\
CIC & 1 & \cite{Al-Alaj2020107} & IEEE Transactions on Network and Service Management & 1 & \cite{10494990} &  & &  \\
CNS & 1 & \cite{10.1109/cns66487.2025.11194984} & IEEE Transactions on Vehicular Technology & 1 & \cite{8528480} &  & &  \\
COMPSAC & 1 & \cite{8377889} & International Journal of Network Management & 1 & \cite{10.1002/nem.1918} &  & &  \\
CPP & 1 & \cite{10.1007/978-3-319-03545-1_3} & Journal of Information Security and Applications & 1 & \cite{DUY2022103080} &  & &  \\
EuroPLoP & 1 & \cite{10.1145/3698322.3698355} & Journal of Network and Systems Management & 1 & \cite{10.1007/s10922-017-9411-6} &  & &  \\
ICCASIT & 1 & \cite{9633403} & Journal of Parallel and Distributed Computing & 1 & \cite{HU2021108} &  & &  \\
ICCCS & 1 & \cite{8888031} & Journal of Reliable Intelligent Environments & 1 & \cite{10.1007/s40860-017-0045-y} &  & &  \\
ICDCN & 1 & \cite{10.1145/3427796.3427798} & Journal of Signal Processing Systems & 1 & \cite{7371471} &  & &  \\
ICNAS & 1 & \cite{10330440} & Mathematical Problems in Engineering & 1 & \cite{10.1155/2017/8740217} &  & &  \\
ICS & 1 & \cite{9359040} & World Wide Web Journal & 1 & \cite{10.1007/s11280-022-01130-2} &  & &  \\
ICST & 1 & \cite{10988999} & Ad Hoc Networks & 1 & \cite{Serrat_2026} &  & &  \\
IM & 1 & \cite{9464034} & Distributed Ledger Technologies: Research and Practice & 1 & \cite{Sinha_2026} &  & &  \\
INFOCOM & 1 & \cite{10229058} & Scientific Reports & 1 & \cite{Shahzad_2026} &  & &  \\
NCA & 1 & \cite{8935066} & WSEAS Transactions on Information Science and Applications & 1 & \cite{Almarhabi_2026} &  & &  \\
NetSys & 1 & \cite{7089082} & Computing & 1 & \cite{Zainal_2026} &  & &  \\
NETWORKS & 1 & \cite{7751150} &  & &  &  & &  \\
NoF & 1 & \cite{9942644} &  & &  &  & &  \\
NOMS & 1 & \cite{9789775} &  & &  &  & &  \\
NSDI & 1 & \cite{10.5555/2616448.2616459} &  & &  &  & &  \\
PDCAT & 1 & \cite{7943369} &  & &  &  & &  \\
PIMRC & 1 & \cite{11275079} &  & &  &  & &  \\
PlatCon & 1 & \cite{8669405} &  & &  &  & &  \\
PLDI & 1 & \cite{10.1145/2594291.2594317} &  & &  &  & &  \\
SACMAT & 1 & \cite{10.1145/2914642.2914647} &  & &  &  & &  \\
SDN and NFV Security & 1 & \cite{10.1007/978-3-319-71761-6_6} &  & &  &  & &  \\
SoICT & 1 & \cite{10.1145/3368926.3369709} &  & &  &  & &  \\
SOSR & 1 & \cite{10.1145/3185467.3185468} &  & &  &  & &  \\
TPS-ISA & 1 & \cite{10.1109/tps-isa50397.2020.00032} &  & &  &  & &  \\
TrustCom & 1 & \cite{Banse2015834} &  & &  &  & &  \\
ICBDS & 1 & \cite{10837050} &  & &  &  & &  \\
WISA & 1 & \cite{10.1007/978-3-030-39303-8_4} &  & &  &  & &  \\
IES & 1 & \cite{Hartono_2026} &  & &  &  & &  \\
\midrule
Overall & 67 & & & 44 & & & 11 & \\
\bottomrule
\end{tabular}
\end{adjustbox}

%% file: Figures/rq1_temporal.tex
\begin{adjustbox}{width=0.85\linewidth, center}
\begin{tikzpicture}
\begin{axis}[
    width=15cm, height=7cm,
    ybar stacked, bar width=0.5cm,
    ylabel={\# of papers},
    ymin=0, ymax=16,
    xtick=data, ytick align=outside,
    xticklabels={2012,2013,2014,2015,2016,2017,2018,2019,2020,2021,2022,2023,2024,2025,2026},
    x tick label style={font=\footnotesize, rotate=0, /pgf/number format/1000 sep={}},
    ytick={0,2,4,6,8,10,12,14,16},
    y tick label style={font=\footnotesize},
    legend style={at={(0.02,0.99)}, anchor=north west, legend columns=3,
        font=\footnotesize, draw=none, legend cell align=left},
    legend image code/.code={\fill[#1] (0cm,-0.1cm) rectangle (0.3cm,0.2cm);},
    ymajorgrids=true, grid style={dotted, gray!25},
    nodes near coords, nodes near coords align={vertical},
    every node near coord/.append style={font=\tiny},
    point meta=explicit symbolic,
    enlarge x limits=0.03,
    axis lines*=left, axis line style={-},
]
\addplot[ybar, fill=color_graph, draw=none] coordinates {
    (0,0) [] (1,3) [3] (2,6) [6] (3,6) [6] (4,4) [4] (5,7) [7] (6,8) [8]
    (7,6) [6] (8,7) [7] (9,5) [5] (10,3) [3] (11,3) [3] (12,3) [3] (13,5) [5] (14,1) [1]
};
\addplot[ybar, fill=color_graph1, draw=none] coordinates {
    (0,0) [] (1,0) [] (2,0) [] (3,0) [] (4,2) [2] (5,4) [4] (6,4) [4]
    (7,8) [8] (8,3) [3] (9,2) [2] (10,2) [2] (11,1) [1] (12,7) [7] (13,4) [4] (14,7) [7]
};
\addplot[ybar, fill=color_graph2, draw=none,
    every node near coord/.append style={font=\tiny, yshift=2pt}] coordinates {
    (0,1) [1] (1,0) [] (2,0) [] (3,0) [] (4,2) [2] (5,3) [3] (6,2) [2]
    (7,1) [1] (8,1) [1] (9,0) [] (10,1) [1] (11,0) [] (12,0) [] (13,0) [] (14,0) []
};
\legend{Conference, Journal, Workshop}
\end{axis}
\end{tikzpicture}
\end{adjustbox}
\caption{Annual distribution of the 122 primary studies by venue type (2012--2026).}
\label{fig:rq1a}

%% file: Figures/sunburst.tex
\begin{figure}[h!]
\centering
\definecolor{redGray2}{RGB}{193,109,97}
\definecolor{redGray1}{RGB}{226,176,168}
\begin{adjustbox}{max width=\textwidth, max height=0.95\textheight, center}
\begin{tikzpicture}[scale=2.1, font=\normalsize]
  \fill[white] (0,0) circle (0.8cm);
  \node[align=center, font=\bfseries\large] at (0,0) {Software \\ Security in SDN\\(122)};
  \fill[blueGray2] (0.00:0.8) -- (0.00:1.9) arc (0.00:212.46:1.9) -- (212.46:0.8) arc (212.46:0.00:0.8) -- cycle;
  \node[align=center, font=\large\bfseries] at (106.23:1.20) {SDN Controller\\(72)};
  \fill[blueGray1!100] (0.00:1.9) -- (0.00:3.9) arc (0.00:41.31:3.9) -- (41.31:1.9) arc (41.31:0.00:1.9) -- cycle;
  \node[align=center, font=\normalsize] at (20.66:3.25) {Access Control\\(14)};
  \node[font=\scriptsize, rotate=3.88] at (3.88:2.62) {\cite{10.1145/3134600.3134603, 7997249, 10.1145/2914642.2914647, 8935066}};
  \node[font=\scriptsize, rotate=12.23] at (12.23:2.62) {\cite{10.1109/tps-isa50397.2020.00032, 8343872, ALIYU2020107421, 8865651, SOUSA2024110130}};
  \node[font=\scriptsize, rotate=20.68] at (20.68:2.62) {\cite{ELZOGHBI2025108169, KANG20191, 10.1002/sec.1369, 10.1145/3309194.3309195, Shahzad_2026}};
  \node[font=\scriptsize, rotate=35.03] at (29.03:2.62) {\cite{10.1145/2660267.2660353, 8422405, 10.1145/3319535.3363214}*};
  \node[font=\scriptsize, rotate=37.38] at (37.38:2.85) {\cite{Porras2015, 10.1145/3301305, 10.1155/2018/9178425, 10.1145/2342441.2342466, Habib202223, 10.1145/3698322.3698355, 7943369}*};
  \fill[blueGray1!90] (41.31:1.9) -- (41.31:3.9) arc (41.31:85.57:3.9) -- (85.57:1.9) arc (85.57:41.31:1.9) -- cycle;
  \node[align=center, font=\normalsize] at (63.44:3.40) {Vulnerability\\Discovery (15)};
  \node[font=\scriptsize] at (63.44:3.05) {\cite{10.1145/3243734.3243799, 10.1007/978-3-319-66332-6_14, 10.1007/978-3-030-39303-8_4, 10988999, Xu2017451}};
  \node[font=\scriptsize] at (63.44:2.85) {\cite{Ujcich2020, Cao2020, 10.1109/TNET.2022.3140824, 10534286, 10.1145/3641541}};
  \node[font=\scriptsize] at (63.44:2.65) {\cite{8718586, 10494990, 8725649, 10.1109/TNET.2017.2748159, Ollando_2026}};
  \node[font=\scriptsize] at (63.44:2.45) {\cite{10.1145/2914642.2914647, 10229058, 9152642, Lee2017, LEE2020101720}*};
  \node[font=\scriptsize] at (63.44:2.25) {\cite{10.1007/s11036-018-1141-9, 10.1002/nem.1918}*};
  \fill[blueGray1!80] (85.57:1.9) -- (85.57:3.9) arc (85.57:123.93:3.9) -- (123.93:1.9) arc (123.93:85.57:1.9) -- cycle;
  \node[align=center, font=\normalsize] at (104.75:3.55) {Mitigation\\(13)};
  \node[font=\scriptsize] at (104.75:3.2) {\cite{10.1145/3698322.3698355, 7943369, 7089082, 10.1145/2660267.2660353, 8422405}};
  \node[font=\scriptsize] at (104.75:3.0) {\cite{Porras2015, 10.1145/3301305, Alsalamh2024, 10.1155/2018/9178425, 10.1002/nem.1918}};
  \node[font=\scriptsize] at (104.75:2.8) {\cite{10.1145/2342441.2342466, Habib202223, 10837050}};
  \node[font=\scriptsize] at (104.75:2.6) {\cite{10.1145/3243734.3243759, 11275079, 9633403, 10.14722/ndss.2015.23283, 7976378}*};
  \node[font=\scriptsize] at (104.75:2.4) {\cite{10.1145/3229616.3229620, 8169856}*};
  \fill[blueGray1!70] (123.93:1.9) -- (123.93:3.9) arc (123.93:153.44:3.9) -- (153.44:1.9) arc (153.44:123.93:1.9) -- cycle;
  \node[align=center, font=\normalsize] at (138.69:3.45) {Security\\Assessment (10)};
  \node[font=\scriptsize] at (138.69:3.0) {\cite{7258233, 10.1007/978-3-319-71761-6_6, 7751150, Kim20234463}};
  \node[font=\scriptsize] at (138.69:2.7) {\cite{10.1145/2684793.2684796, 8669405, Lee2017}};
  \node[font=\scriptsize] at (138.69:2.4) {\cite{10.1007/s40860-017-0045-y, LEE2020101720, 10.1155/2017/8740217}};
  \node[font=\scriptsize] at (138.69:2.1) {\cite{10.1145/3658644.3670301}*};
  \fill[blueGray1!60] (153.44:1.9) -- (153.44:3.9) arc (153.44:177.05:3.9) -- (177.05:1.9) arc (177.05:153.44:1.9) -- cycle;
  \node[align=center, font=\normalsize] at (165.25:3.38) {Exploitation\\(8)};
  \node[font=\scriptsize, rotate=336.31] at (156.31:2.60) {\cite{8023155, 10.1145/3243734.3243759, 10229058}};
  \node[font=\scriptsize, rotate=339.92] at (159.92:2.60) {\cite{10.1007/978-3-319-26362-5_16, 9152642, 10.1145/3658644.3670301}};
  \node[font=\scriptsize, rotate=343.52] at (163.52:2.60) {\cite{10.14722/ndss.2015.23283, 10.1145/2876019.2876024}};
  \node[font=\scriptsize, rotate=347.13] at (167.13:2.60) {\cite{10.1145/3243734.3243799, 10.1007/978-3-319-66332-6_14, Xu2017451}*};
  \node[font=\scriptsize, rotate=350.74] at (170.74:2.60) {\cite{Ujcich2020, Cao2020, 8718586}*};
  \node[font=\scriptsize, rotate=354.34] at (174.34:2.60) {\cite{10494990, 10.1002/sec.1369}*};
  \fill[blueGray1!50] (177.05:1.9) -- (177.05:3.9) arc (177.05:194.75:3.9) -- (194.75:1.9) arc (194.75:177.05:1.9) -- cycle;
  \node[align=center, font=\normalsize] at (185.90:3.35) {Malicious\\Behavior Det.\\(6)};
  \node[font=\scriptsize, rotate=0.94] at (180.94:2.55) {\cite{10.1007/978-3-030-19086-6_60, 11275079, DAM2025104677}};
  \node[font=\scriptsize, rotate=5.90] at (185.90:2.55) {\cite{10.1145/3229616.3229620, 8169856, Liu_2025}};
  \node[font=\scriptsize, rotate=10.86] at (190.86:2.55) {\cite{Shahzad_2026}*};
  \fill[blueGray1!40] (194.75:1.9) -- (194.75:3.9) arc (194.75:209.51:3.9) -- (209.51:1.9) arc (209.51:194.75:1.9) -- cycle;
  \node[align=center, font=\normalsize] at (202.13:3.45) {Verification\\(5)};
  \node[font=\scriptsize, rotate=18.52] at (198.52:2.55) {\cite{10.1145/3319535.3363214, 9633403, 10.1145/2594291.2594317}};
  \node[font=\scriptsize, rotate=25.75] at (205.75:2.55) {\cite{7976378, 10.1007/s11036-018-1141-9}};
  \fill[blueGray1!30] (209.51:1.9) -- (209.51:3.9) arc (209.51:212.46:3.9) -- (212.46:1.9) arc (212.46:209.51:1.9) -- cycle;
  \node[align=center, font=\normalsize, rotate=30.98] at (210.98:3.50) {RCA (1)};
  \node[font=\scriptsize, rotate=30.98] at (210.98:2.55) {\cite{272212}};
  \fill[green2] (212.46:0.8) -- (212.46:1.9) arc (212.46:283.28:1.9) -- (283.28:0.8) arc (283.28:212.46:0.8) -- cycle;
  \node[align=center, font=\large\bfseries] at (247.87:1.35) {Interfaces\\(24)};
  \fill[green1!100] (212.46:1.9) -- (212.46:3.9) arc (212.46:268.52:3.9) -- (268.52:1.9) arc (268.52:212.46:1.9) -- cycle;
  \node[align=center, font=\normalsize] at (240.49:3.35) {Access Control\\(19)};
  \node[font=\scriptsize] at (240.49:3.05) {\cite{10.1145/2486001.2486003, 10.1145/3368926.3369709, 7997245, 6980437, 7579735, Banse2015834, LENG201968}};
  \node[font=\scriptsize] at (240.49:2.85) {\cite{DUY2022103080, 8528480, 10.1007/s11280-022-01130-2, 9452113, 7371471, PALADI2019155, 10.1007/s12243-025-01133-w}};
  \node[font=\scriptsize] at (240.49:2.65) {\cite{HU2021108, 10.1007/s10922-017-9411-6, 9162669, Almarhabi_2026, Zainal_2026}};
  \fill[green1!90] (268.52:1.9) -- (268.52:3.9) arc (268.52:274.43:3.9) -- (274.43:1.9) arc (274.43:268.52:1.9) -- cycle;
  \node[align=center, font=\normalsize, rotate=271.48] at (271.48:3.35) {Mitigation (2)};
  \node[font=\scriptsize, rotate=271.48] at (271.48:2.65) {\cite{9942644, Hartono_2026}};
  \node[font=\scriptsize, rotate=271.48] at (271.48:2.20) {\cite{DUY2022103080, 8528480, Serrat_2026}*};
  \fill[green1!80] (274.43:1.9) -- (274.43:3.9) arc (274.43:277.38:3.9) -- (277.38:1.9) arc (277.38:274.43:1.9) -- cycle;
  \node[align=center, font=\normalsize, rotate=275.90] at (275.90:3.30) {Verification (1)};
  \node[font=\scriptsize, rotate=275.90] at (275.90:2.55) {\cite{8377889}};
  \fill[green1!70] (277.38:1.9) -- (277.38:3.9) arc (277.38:280.33:3.9) -- (280.33:1.9) arc (280.33:277.38:1.9) -- cycle;
  \node[align=center, font=\normalsize, rotate=278.85] at (278.85:3.30) {Vuln. Disc. (1)};
  \node[font=\scriptsize, rotate=278.85] at (278.85:2.60) {\cite{6733671}};
  \node[font=\scriptsize, rotate=278.85] at (278.85:2.22) {\cite{8725649, 8377889}*};
  \fill[green1!60] (280.33:1.9) -- (280.33:3.9) arc (280.33:283.28:3.9) -- (283.28:1.9) arc (283.28:280.33:1.9) -- cycle;
  \node[align=center, font=\normalsize, rotate=281.80] at (281.80:3.25) {Exploitation (1)};
  \node[font=\scriptsize, rotate=281.80] at (281.80:2.55) {\cite{Serrat_2026}};
  \fill[orange2] (283.28:0.8) -- (283.28:1.9) arc (283.28:321.64:1.9) -- (321.64:0.8) arc (321.64:283.28:0.8) -- cycle;
  \node[align=center, font=\large\bfseries] at (302.46:1.45) {SDN\\Applications\\(13)};
  \fill[orange1!100] (283.28:1.9) -- (283.28:3.9) arc (283.28:292.13:3.9) -- (292.13:1.9) arc (292.13:283.28:1.9) -- cycle;
  \node[align=center, font=\normalsize, rotate=287.70] at (287.70:3.3) {Verification (3)};
  \node[font=\scriptsize, rotate=287.70] at (287.70:2.32) {\cite{10.1007/978-3-319-03545-1_3, YAO2025111259, 8109275}};
  \fill[orange1!90] (292.13:1.9) -- (292.13:3.9) arc (292.13:300.98:3.9) -- (300.98:1.9) arc (300.98:292.13:1.9) -- cycle;
  \node[align=center, font=\normalsize, rotate=296.56] at (296.56:3.12) {Mal. Behav. Det. (3)};
  \node[font=\scriptsize, rotate=296.56] at (296.56:2.15) {\cite{8526819, 9359040, Sinha_2026}};
  \fill[orange1!80] (300.98:1.9) -- (300.98:3.9) arc (300.98:306.89:3.9) -- (306.89:1.9) arc (306.89:300.98:1.9) -- cycle;
  \node[align=center, font=\normalsize, rotate=303.93] at (303.93:3.3) {Access Ctrl (2)};
  \node[font=\scriptsize, rotate=303.93] at (303.93:2.55) {\cite{8888031, Al-Alaj2020107}};
  \node[font=\scriptsize, rotate=303.93] at (303.93:2.25) {\cite{ALIYU2020107421}*};
  \fill[orange1!70] (306.89:1.9) -- (306.89:3.9) arc (306.89:312.79:3.9) -- (312.79:1.9) arc (312.79:306.89:1.9) -- cycle;
  \node[align=center, font=\normalsize, rotate=309.84] at (309.84:3.3) {Sec. Assess. (2)};
  \node[font=\scriptsize, rotate=309.84] at (309.84:2.45) {\cite{10.1145/2684793.2684797, 10.1145/2876019.2876026}};
  \fill[orange1!60] (312.79:1.9) -- (312.79:3.9) arc (312.79:315.74:3.9) -- (315.74:1.9) arc (315.74:312.79:1.9) -- cycle;
  \node[align=center, font=\normalsize, rotate=314.26] at (314.26:3.3) {Vuln. Disc. (1)};
  \node[font=\scriptsize, rotate=314.26] at (314.26:2.55) {\cite{10330440}};
  \fill[orange1!50] (315.74:1.9) -- (315.74:3.9) arc (315.74:318.69:3.9) -- (318.69:1.9) arc (318.69:315.74:1.9) -- cycle;
  \node[align=center, font=\normalsize, rotate=317.21] at (317.21:3.3) {Mitigation (1)};
  \node[font=\scriptsize, rotate=317.21] at (317.21:2.55) {\cite{9789775}};
  \node[font=\scriptsize, rotate=317.21] at (317.21:2.25) {\cite{10.1145/2660267.2660353}*};
  \fill[orange1!40] (318.69:1.9) -- (318.69:3.9) arc (318.69:321.64:3.9) -- (321.64:1.9) arc (321.64:318.69:1.9) -- cycle;
  \node[align=center, font=\normalsize, rotate=320.16] at (320.16:3.25) {Exploitation (1)};
  \node[font=\scriptsize, rotate=320.16] at (320.16:2.55) {\cite{10.1007/s11036-023-02156-0}};
  \node[font=\scriptsize, rotate=320.16] at (320.16:2.2) {\cite{9789775, 10.1145/3243734.3243759}*};
  \fill[redGray2] (321.64:0.8) -- (321.64:1.9) arc (321.64:360.00:1.9) -- (360.00:0.8) arc (360.00:321.64:0.8) -- cycle;
  \node[align=center, font=\large\bfseries] at (340.82:1.38) {Data Plane\\(13)};
  \fill[redGray1!100] (321.64:1.9) -- (321.64:3.9) arc (321.64:339.34:3.9) -- (339.34:1.9) arc (339.34:321.64:1.9) -- cycle;
  \node[align=center, font=\normalsize] at (330.49:3.45) {Exploitation\\(6)};
  \node[font=\scriptsize, rotate=325.53] at (325.53:2.58) {\cite{10.1145/3185467.3185468, 9464034, 10.1109/netsoft54395.2022.9844121}};
  \node[font=\scriptsize, rotate=330.49] at (330.49:2.58) {\cite{10.1145/3445814.3446764, 10.1109/cns66487.2025.11194984, 10.1145/3140649.3140651}};
  \node[font=\scriptsize, rotate=335.45] at (335.45:2.58) {\cite{10.1109/sp61157.2025.00194, 8428482}*};
  \fill[redGray1!90] (339.34:1.9) -- (339.34:3.9) arc (339.34:345.25:3.9) -- (345.25:1.9) arc (345.25:339.34:1.9) -- cycle;
  \node[align=center, font=\normalsize, rotate=342.30] at (342.30:3.3) {Vuln. Disc. (2)};
  \node[font=\scriptsize, rotate=342.30] at (342.30:2.55) {\cite{10.1109/sp61157.2025.00194, 10.1145/3427796.3427798}};
  \node[font=\scriptsize, rotate=342.30] at (342.30:2.25) {\cite{10.5555/2616448.2616459, 10.1145/3140649.3140651}*};
  \fill[redGray1!80] (345.25:1.9) -- (345.25:3.9) arc (345.25:351.15:3.9) -- (351.15:1.9) arc (351.15:345.25:1.9) -- cycle;
  \node[align=center, font=\normalsize, rotate=348.20] at (348.20:3.28) {Verification (2)};
  \node[font=\scriptsize, rotate=348.20] at (348.20:2.4) {\cite{10.1145/3230543.3230548, 10.5555/2616448.2616459}};
  \fill[redGray1!70] (351.15:1.9) -- (351.15:3.9) arc (351.15:360.00:3.9) -- (360.00:1.9) arc (360.00:351.15:1.9) -- cycle;
  \node[align=center, font=\normalsize, rotate=355.57] at (355.57:3.3) {Mitigation (3)};
  \node[font=\scriptsize, rotate=355.57] at (355.57:2.52) {\cite{8428482, 10.1109/tdsc.2023.3277939, Su_2026}};
  \node[font=\scriptsize, rotate=355.57] at (355.57:2.1) {\cite{10.1145/3185467.3185468}*};
  \foreach \a in {212.46, 283.28, 321.64} {\draw[white, line width=1pt] (\a:0.8) -- (\a:1.9);}
  \foreach \a in {41.31, 85.57, 123.93, 153.44, 177.05, 194.75, 209.51, 212.46, 268.52, 274.43, 277.38, 280.33, 283.28, 292.13, 300.98, 306.89, 312.79, 315.74, 318.69, 321.64, 339.34, 345.25, 351.15} {\draw[white, line width=0.8pt] (\a:1.9) -- (\a:3.9);}
  \draw[white, line width=2pt] (0,0) circle (0.8);
  \draw[white, line width=1.5pt] (0,0) circle (1.9);
  \draw[white, line width=1.5pt] (0,0) circle (3.9);
\end{tikzpicture}
\end{adjustbox}
\caption{\textbf{Classification of SDN software security research themes (RQ1), illustrated as a sunburst: inner ring = SDN software component, outer ring = security theme. Bracketed numbers are study counts; leaf nodes list the corresponding studies. Entries marked with an asterisk (*) indicate a secondary classification of a study primarily classified elsewhere.}}
\label{fig:slr_rq1_topics}
\end{figure}

%% file: Sections/9-challenges.tex
We derive five challenges from the results. Each starts with a finding from this review, adds the limitations that the primary studies themselves report, and ends with the directions that follow.

\noindent
\textbf{Defenses for programmable and virtual switches.} The corpus contains no formal verification of data-plane software after Vera in 2018, and Vera finds programming bugs without an attacker model~\cite{10.1145/3230543.3230548}. The weakness class missing integrity check of loaded programs (RQ3) has no defense in the corpus beyond obfuscation and the countermeasures that the attack studies themselves propose (Table~\ref{tab:dataplane-profiles}). The primary studies report three limits. Extending probe-based defenses to data-plane state explodes the state space to cover, and replicating state across switches may add processing time to every packet~\cite{9464034}. Whether a hardware root of trust on switches can verify the integrity of P4 programs remains an open question~\cite{9464034}. Software mitigations of the Open vSwitch parser defects cost little performance, yet they are rarely considered when new data-plane components are evaluated~\cite{10.1145/3185467.3185468}. The open directions are integrity checking of loaded programs and table entries, verification against an attacker who controls a P4-programmable device, whose capabilities current adversarial verification does not address~\cite{9464034}, and adversarial testing of eBPF programs, for which P4wn has no counterpart~\cite{10.1145/3445814.3446764}.

\noindent
\textbf{Security of third-party applications.} Dynamic testing, the most active family, is absent from the 37 studies of applications and interfaces (Table~\ref{tab:rq2-matrix}), although access control, which governs which applications may invoke controller resources, is the most studied concern, with 35 studies (Section~\ref{sec:rq1result}). Most access-control schemes are built on a single controller, 15 of the 35 on Floodlight. Most controllers provide no mechanism to authenticate the developer or provenance of an application (RQ3)~\cite{ALIYU2020107421}, although application and vendor certificates were proposed in 2015~\cite{Banse2015834}, and isolating each application in its own process, as ROSEMARY does~\cite{10.1145/2660267.2660353}, would require a significant redesign of a controller such as ONOS~\cite{10.1145/3243734.3243759}. The authors of VOGUE found that existing permission systems are tightly coupled with a specific controller and cannot effectively defeat attack trials~\cite{8865651}, and cross-app poisoning escapes access control that does not track information flow between applications (RQ4). Permission systems should be tested by running malicious applications against them, as VOGUE does with known attacks~\cite{8865651}, on a common set such as the malicious applications that INDAGO collected from earlier SDN security projects~\cite{8526819}. Permission models decoupled from the controller~\cite{8865651}, information-flow control~\cite{10.1145/3243734.3243759}, and controller designs that authenticate and isolate applications would address these limits.

\noindent
\textbf{Analysis across components.} The taxonomy organizes weaknesses by component, but some defects belong to no single plane (RQ3), and cross-plane techniques appeared only in 2020 for event analysis and in 2025 for fuzzing (Section~\ref{sec:rq2result}). Few studies analyze these interactions, through event-flow analysis~\cite{Ujcich2020}, data-dependency analysis~\cite{9152642}, information-flow control~\cite{10.1145/3243734.3243759, 9789775}, and multi-plane fuzzing~\cite{10.1109/sp61157.2025.00194}. Fuzzers that target one plane do not consider, or simplify, the interactions with the other plane~\cite{10.1109/sp61157.2025.00194}, and information-flow control in ProvSDN assumes a single controller and a single switch pipeline~\cite{9789775}. Information-flow control should extend to several controllers and virtualized pipelines, for which vIFC is a conceptual solution~\cite{9789775}. Static and dynamic analysis should be combined, as the authors of the data-dependency analysis and of INDAGO propose~\cite{9152642, 8526819}. The implementations of EventScope, the data-dependency analysis, vIFC, and Chimera are publicly available.

\noindent
\textbf{Runtime detection and forensics.} Ten studies detect malicious behavior or analyze attacks after the fact, against 35 on access control (Section~\ref{sec:rq1result}). None of the ten targets the data plane, although edits to table entries can alter stateful behavior without appearing in the logs of ONOS~\cite{9464034}, and only PicoSDN among them makes its implementation available. Five studies of the corpus build on data provenance~\cite{10.1145/3243734.3243759, 9789775, 272212, DAM2025104677, Liu_2025}. Recording provenance adds 7 to 21~ms to each new forwarding rule in PicoSDN~\cite{272212}, and reading it from the controller logs means trusting the controller that writes them~\cite{DAM2025104677}. Provenance should be collected inside the controller rather than from its logs and should not depend on trust in the controller~\cite{DAM2025104677}, and detection should extend to the data plane.

\noindent
\textbf{Evaluation and comparison.} Comparison with earlier techniques or tools has been common only since 2024, and few studies make their artifact available (Table~\ref{tab:research-practice}). Controller fuzzers compare with DELTA and BEADS (Section~\ref{sec:rq1result}), whose implementations are public~\cite{Lee2017, 10.1007/978-3-319-66332-6_14}. Two studies evaluate on a university datacenter and on an SD-WAN that spans two campus networks and one enterprise network, and both found vulnerabilities that received CVE identifiers~\cite{10.1145/3243734.3243799, 10534286}. Later work should release its implementations together with the attacks used to evaluate them, so that others can compare with them, and should evaluate on such networks.

%% file: Sections/6-threatsToValidity.tex
Like any systematic review, ours faces threats to validity, which we discuss here together with the steps we took to reduce them.

\noindent
\textbf{Construct and coverage validity.} Our review may not be exhaustive. We searched four major repositories, namely IEEE Xplore, the ACM Digital Library, SpringerLink, and ScienceDirect, chosen for their broad coverage and frequent use in prior reviews~\cite{LI201767, 10.1145/3073565}, but restricting the search to these databases and to a fixed set of terms may have missed relevant studies. To reduce this risk we paired the database search with backward and forward snowballing and screened with high recall, carrying doubtful studies forward to full-text assessment rather than excluding them early. The snowballing step alone added 31 studies that the keyword search had not surfaced, which gives some evidence that the combined strategy recovers work the automated search misses.

\noindent
\textbf{Internal validity.} Study selection and quality assessment can introduce bias. We mitigated this by fixing the inclusion and exclusion criteria before screening, by assessing every retained study against four explicit quality-assessment criteria, and by keeping only those above a predefined threshold. The full criteria, scores, and decisions are in our replication package, so the selection can be inspected and reproduced.

\noindent
\textbf{Conclusion validity.} Classifying the studies and building the vulnerability and attack taxonomies involves subjective judgment. We reduced this threat through open coding, assigning codes to the source text and refining the categories iteratively as the analysis progressed, and we recorded the resulting codebook in the replication package so that the coding can be audited.

\noindent
\textbf{External validity.} Our findings describe the published literature on SDN software security and should be read as such. They characterize where research effort has concentrated and where it has not, rather than the absolute prevalence of vulnerabilities in deployed SDN systems, and readers should interpret the gaps we identify as gaps in the literature that motivate future study.

%% file: Sections/7-relatedWork.tex
Security in software-defined networking has been surveyed extensively, but from a perspective that differs from ours. Existing reviews organize the field by SDN architectural plane (application, control, and data) and frame security in network terms. They catalog attacks such as denial of service, spoofing, tampering, man-in-the-middle and ARP poisoning, link flooding, side and covert-channel leakage, and flow-rule or topology poisoning. The countermeasures they review are likewise network-level: authentication and authorization, channel encryption, cryptographic and blockchain protocols, and entropy or learning-based detection~\cite{9527257, 10.1007/s40860-022-00171-8, 10.1007/s10776-022-00561-y, 10.1145/3605801.3605814, 9988945}. Several structure this coverage with threat-modeling frameworks such as STRIDE or the ITU-T security dimensions~\cite{9527257, 7226783}.

Three sub-directions recur within this body of work. Control-plane reviews treat the controller as a centralized choke point and a single point of failure, and focus on denial of service, spoofing, and authentication or secure-communication defenses~\cite{10226193, 11292437, 10.1002/cpe.5300}. Data-plane reviews examine attacks on switches and the southbound interface, such as state exhaustion in stateful pipelines, packet injection, and topology manipulation, together with the defenses against them~\cite{7890396, 10.1145/3605801.3605814}. Broader reviews fold both planes and the interfaces into a single layered taxonomy~\cite{9527257, 10.1007/s40860-022-00171-8, 9760465}, and some specialize further, for instance in low-rate, stealthy attacks across programmable networks~\cite{10.1145/3704434}.

A few of these reviews are themselves systematic, and are the closest to ours in method. One systematic literature review catalogs SDN threats, their causes, and per-plane solutions, with a distinctive emphasis on the deployment cost and the software and hardware requirements of those solutions~\cite{10.1007/s10776-022-00561-y}; others apply a systematic protocol to per-layer threats and mitigations~\cite{10.1007/s40860-022-00171-8} or to control-plane authentication and encryption~\cite{11292437}. Their methodology is comparable to ours, but their object of study remains the network.

What these surveys do not provide is a treatment of SDN components as software. Controllers, applications, interfaces, and data-plane logic are analyzed as network entities whose compromise disrupts the network, rather than as software artifacts whose code, implementation, or logic can contain exploitable defects. Where software concerns do surface, they appear only incidentally, in a passing mention of malicious-application detection through behavior graphs or control-flow analysis~\cite{9527257}, of conceptual vulnerability scenarios for stateful data planes~\cite{7890396}, or of hardening the controller's operating system~\cite{10.1002/cpe.5300}. None of these surveys examines the software-engineering methods used to find such defects, namely fuzzing, static and dynamic analysis, symbolic execution, model checking, and formal verification.

The work closest to our software-engineering perspective lies outside the security surveys. An empirical study analyzes more than 500 critical bugs across three widely used SDN controllers and organizes them into a taxonomy~\cite{9505089}, showing that the control plane, as software, is bug-prone and amenable to systematic defect analysis. That study, however, targets dependability (fault detection and recovery) rather than security, and it mines controller code directly rather than reviewing the literature. Its existence underlines the gap we address: the software-defect lens has been applied to SDN reliability, but not, systematically, to SDN software security.

Our review addresses this gap. To our knowledge, it is the first systematic literature review to study SDN security from a software-engineering standpoint. It organizes the literature by software component, namely the controller, applications, interfaces, and data plane, and by the security activity applied to each, it analyzes the methods used to test and analyze SDN software for defects (RQ2), and it derives taxonomies of software vulnerabilities (RQ3) and of the attacks that exploit them (RQ4). It thus occupies the space the prior literature leaves open, treating the software security of SDN as a subject for systematic review in its own right. Table~\ref{tab:relwork} summarizes how our review differs from prior reviews along four dimensions.

\begin{table}[ht]
\caption{Prior SDN-security reviews compared with this SLR. ``SE focus'' denotes a software-engineering software-security focus (\textcolor{rd}{\ding{55}}~none, $\sim$~incidental, \textcolor{rq}{\ding{51}}~primary).}
\label{tab:relwork}
\centering
\begin{adjustbox}{width=\linewidth,center}
\begin{tabular}{l l l c c}
\toprule
\textbf{Review (year)} & \textbf{Primary focus} & \textbf{Components} & \textbf{SE focus} & \textbf{Systematic} \\
\midrule
Control-plane (in)security~\cite{10226193} (2023) & Control-plane attacks (DoS/DDoS) & Control plane & \textcolor{rd}{\ding{55}} & \textcolor{rd}{\ding{55}} \\
Control-plane approaches~\cite{11292437} (2025) & Authentication \& secure communication & Control plane & \textcolor{rd}{\ding{55}} & \textcolor{rq}{\ding{51}} \\
Next-generation controllers~\cite{10.1002/cpe.5300} (2019) & Controller-targeted attacks & Control plane & $\sim$ & \textcolor{rd}{\ding{55}} \\
Stateful data planes~\cite{7890396} (2017) & State exhaustion / DoS & Data plane & $\sim$ & \textcolor{rd}{\ding{55}} \\
Data-plane attacks \& defense~\cite{10.1145/3605801.3605814} (2023) & Switch and southbound-API attacks & Data plane & \textcolor{rd}{\ding{55}} & \textcolor{rd}{\ding{55}} \\
Low-rate threats~\cite{10.1145/3704434} (2024) & Low-rate / stealthy attacks & All planes + PDP & $\sim$ & \textcolor{rd}{\ding{55}} \\
Main issues \& solutions~\cite{9527257} (2021) & STRIDE, layer by layer & All planes + interfaces & $\sim$ & \textcolor{rd}{\ding{55}} \\
Security in SDN~\cite{7226783} (2015) & Per-plane threats; ITU-T & App/control/data & $\sim$ & \textcolor{rd}{\ding{55}} \\
Threats, solutions, applications~\cite{9988945} (2022) & SDN as security tool vs.\ threats & All planes & \textcolor{rd}{\ding{55}} & \textcolor{rd}{\ding{55}} \\
Comprehensive SDN security~\cite{10.1007/s40860-022-00171-8} (2023) & Per-layer threats / mitigations & App/control/data & $\sim$ & \textcolor{rq}{\ding{51}} \\
Threat-taxonomy review~\cite{9760465} (2022) & Per-plane + cross-layer threats & Control/data/app & $\sim$ & \textcolor{rd}{\ding{55}} \\
Privacy \& security SLR~\cite{10.1007/s10776-022-00561-y} (2022) & Threats + solution cost & App/control/data + APIs & \textcolor{rd}{\ding{55}} & \textcolor{rq}{\ding{51}} \\
\midrule
Bugs in SDN~\cite{9505089} (2021) & Controller bugs, fault-tolerance & Control plane & reliability\textsuperscript{*} & \textcolor{rd}{\ding{55}} \\
\midrule
\textbf{This SLR} & \textbf{Software-engineering security of SDN} & \textbf{All software components} & \textbf{\textcolor{rq}{\ding{51}}} & \textbf{\textcolor{rq}{\ding{51}}} \\

\bottomrule
\end{tabular}
\end{adjustbox}

\smallskip
{\footnotesize \textsuperscript{*}Treats SDN as software, but targets reliability/fault-tolerance, not security.}
\end{table}

%% file: Sections/8-conclusion.tex
We have presented a systematic literature review of 122 primary studies, published between 2012 and 2026, on the security of the software that constitutes software-defined networking. Where prior surveys treat SDN security as a networking problem, we treat the controller, the applications, the interfaces, and the data plane as software artifacts whose code, logic, and interactions can be defective, and we organize the literature accordingly. The review answers four questions. It charts how the field has grown, matured into journals, and shifted from securing SDN by design toward discovering vulnerabilities in deployed implementations (RQ1). It characterizes the four families of technique used to find software defects, namely dynamic testing, static analysis, formal methods, and learning-based detection, and shows which components each family reaches (RQ2). It derives a component-organized taxonomy of vulnerabilities (RQ3) and a complementary taxonomy of attacks that maps each pattern to the weaknesses it exploits and the defenses reported against it (RQ4).

Three findings stand out. The controller absorbs most of the research effort while the application layer and the data plane remain comparatively neglected, even though both run privileged, attacker-influenced code. This concentration follows the earlier maturation and the greater accessibility of controller implementations. Comparison with earlier tools, reported by 3 of 24 studies before 2017 and by 14 of 27 since 2024, suggests a sub-field that is leaving its exploratory phase, although publicly available artifacts remain rare. Many of the reported weaknesses are design and logic flaws, particularly missing authorization and unsafe component interaction, rather than memory-safety bugs, which means that techniques able to reason across components, such as information-flow and event-flow analysis, are disproportionately valuable. Defenses are mature on the application and interface side, where permission models, role-based access control, sandboxing, and channel encryption are well developed, but they are thin for the programmable data plane, whose attacks are growing in sophistication.

These observations point to a clear agenda (Section~\ref{sec:challenges}). The community needs analysis and testing tools that reason about whole-system interaction rather than single components, principled isolation and authorization for third-party applications, and defenses for programmable and virtual switches that match the maturity of controller hardening. Runtime detection and forensic root-cause analysis, today represented by only a handful of studies, deserve greater attention as SDN moves into production at scale. We hope that the taxonomies and the consolidated view offered here give researchers and practitioners a foundation on which to build that work, and we release our artifacts to support its replication and extension.